\documentclass[showpacs,superscriptaddress,amsmath,amssymb]{revtex4}
\usepackage{epsfig}

\usepackage{dsfont}

\newcommand{\integer}{\relax{\rm I\kern-.18em N}}

\newcommand{\addrMOSKAU}{Institute for Theoretical and Experimental Physics, B.
Cheremushkinskaya 25, 117259 Moscow, Russia}
\newcommand{\addrTUEBINGEN}{Institut f\"{u}r Theoretische Physik, T\"{u}bingen
Universit\"{a}t, Auf der Morgenstelle 14, D-72076 T\"{u}bingen, Germany}
\begin{document}

\title{Semiclassical expansion of quantum characteristics \\for many-body potential scattering problem}

\author{M. I. Krivoruchenko}
\affiliation{\addrMOSKAU}
\affiliation{\addrTUEBINGEN}

\author{C. Fuchs}
\affiliation{\addrTUEBINGEN}

\author{Amand Faessler}
\affiliation{\addrTUEBINGEN}

\begin{abstract}

In quantum mechanics, systems can be described in phase space in terms of the Wigner function and the star-product operation. Quantum characteristics, which appear in the Heisenberg picture as the Weyl's symbols of operators of canonical coordinates and momenta, can be used to solve the evolution equations for symbols of other operators acting in the Hilbert space. To any fixed order in the Planck's constant, many-body potential scattering problem simplifies to a statistical-mechanical problem of computing an ensemble of quantum characteristics and their derivatives with respect to the initial canonical  coordinates and momenta. The reduction to a system of ordinary differential equations pertains rigorously at any fixed order in $\hbar$. We present semiclassical expansion of quantum characteristics for many-body scattering problem and provide tools for calculation of average values of time-dependent physical observables and cross sections. The method of quantum characteristics admits the consistent incorporation of specific quantum effects, such as non-locality and coherence in propagation of particles, into the semiclassical transport models. We formulate the principle of stationary action for quantum Hamilton's equations and give quantum-mechanical extensions of the Liouville theorem on the conservation of phase-space volume and the Poincar\'e theorem on the conservation of $2p$ forms. The lowest order quantum corrections to the Kepler periodic orbits are constructed. These corrections show the resonance behavior.

\end{abstract}


\maketitle

\section{Introduction}
\setcounter{equation}{0} 

The deformation quantization \cite{WEYL1,WEYL2,WIGNER,GROE,MOYAL,BARLE} has been the focus of renewed interest last decades. It uses the Weyl's association rule \cite{WEYL1,WEYL2} to establish the one-to-one correspondence between phase-space functions and operators in the Hilbert space. The Wigner function \cite{WIGNER} appears as the Weyl's symbol of the density matrix. Refined formulation of the Weyl's association rule is given by Stratonovich \cite{STRA}. 
Groenewold \cite{GROE} proposed the quantum-mechanical description of the evolution in phase space and introduced into the formalism the concept of star-product of phase-space functions. The quantum evolution is determined by the 
skew-symmetric part of the star-product, known as the sine bracket or the Moyal bracket \cite{MOYAL,BARLE}. The Moyal bracket represents the quantum deformation of the Poisson bracket, being essentially unique \cite{MEHTA}. The phase-space quantum dynamics keeps many features of the classical Hamiltonian dynamics.

The formulation of quantum mechanics in phase space and the star-product are reviewed in Refs.
\cite{VOROS,BERRY,BAYEN,CARRU,HILL,BALAZ,ZACHO}. The 
Stratonovich version \cite{STRA} of the Weyl's quantization and dequantization is discussed in Refs. \cite{BALAZ,GRAC-1,GRAC-2,KRF,KRFA,MIKR}. 
Wigner functions have found numerous applications in quantum many-body physics, kinetic theory \cite{ZUBA1,ZUBA2}, collision theory, and quantum chemistry \cite{EVANS,FRENK}. Transport models created initially for needs of quantum chemistry in order to describe chemical reactions were modified and extended for modelling heavy-ion collisions 
\cite{CARRU,BOTE,RQMD,UrQMD,AICHE,TUEB2,TUEB3,TUEB4,rbuu,cassing99}.

First attempts to exploit specific properties inherent to the $\star$-calculus have recently been made towards calculation of higher order terms of the $\hbar$-expansion of the Bohr-Sommerfeld quantization rules \cite{CARGO} and determinants of operators with occur in quantum field theory at one loop \cite{PLETN,BANIN,LOIK}. A diagrammatic method for calculation of symbols of operator functions in form of the series expansion in $\hbar$ has been proposed \cite{OSBO,GRACIA}.

Transport models in heavy-ion physics are designed for a phenomenological description of complicated reaction dynamics of nuclear collisions. There exist
 several well established transport models such as the Boltzmann-Uehling-Uhlenbeck (BUU) approach \cite{cassing99,rbuu}, (Relativistic) Quantum Molecular Dynamics (R)QMD \cite{RQMD,UrQMD,AICHE,TUEB2,TUEB3,TUEB4} or Anti-symmetrized Molecular Dynamics (AMD) \cite{FELD,amd}. These approaches can be derived from quantum field theory
\cite{dani84,BOTE,henning95,knoll98} in semiclassical limit and contain basic quantum features such as final state Pauli blocking for binary collisions of fermions. Numerical solutions are realized by propagating test-particles (BUU) or centroids of wave packets (QMD, AMD) along classical trajectories in phase space. In the case of AMD, wave packets are anti-symmeterized according to their parameters. Transport models provide a sound phenomenological basis for description of a variety of complicated nuclear phenomena. Quantum interference effects are, however, beyond the scope of these models. The problem of intrinsic consistency of approximations remains under the discussion and stimulates further developments (see e.g. \cite{FELD,KOHL,CAJU} and references therein).

The most striking feature of the semiclassical transport models is obviously the
phase-space trajectories along which the particles or wave packets of the particles are supposed to propagate. While scattering of $n$ classical particles can be processed by means of conventional computer programmes, the evolution of many-body wave functions represents a field-theoretic problem that can currently be approached neither analytically nor numerically. Any implementation of the consistent quantum dynamics should obviously keep trajectories as an attribute allowing to access many-body scattering problems numerically.

The notion of phase-space trajectories arises naturally in the deformation quantization through the Weyl's transform of Heisenberg operators of canonical coordinates and momenta \cite{OSBO,MCQUA,KRFA,DIPR}. These trajectories obey the Hamilton's equations in the quantum form and play the role of quantum characteristics in terms of which the time-dependent symbols of operators are expressed. In classical limit, quantum characteristics turn to classical trajectories. The knowledge of the quantum phase flow is equivalent to the knowledge of the quantum dynamics. 

In this work, we report the semiclassical expansion of quantum characteristics for many-body scattering problem, based on the method \cite{OSBO,MCQUA,KRFA,DIPR}, and provide tools for calculation of average values of time-dependent physical observables and scattering cross sections.

We show that to any fixed order in $\hbar$ it is sufficient to work with quantum characteristics, provided equations of motion and rules for calculation of the time dependent average values of physical observables are modified. The quantum evolution problem becomes thereby entirely identical with a statistical-mechanical problem. We hope that, to a fixed order in $\hbar$, the method of quantum characteristics captures basic quantum properties of many-body potential scattering being numerically effective the same time.

We thus propose a self-consistent non-relativistic quantum mechanical approach for solving many-body potential scattering problem of spin-zero particles. The method is based on semiclassical expansion of quantum characteristics and star-functions of quantum characteristics in a power series over the Planck's constant. The evolution equations represent a coupled system of first-order ordinary differential equations (ODE) for quantum trajectories in the phase space and for associated Jacobi fields.

The outline of the paper is as follows: In the next Sect., we overview the Weyl's association rule and the star-product using the method proposed by Stratonovich \cite{STRA}. Sect. III is devoted to discussion of quantum characteristics. We investigate transformation properties of canonical variables and phase-space functions under unitary transformations in the Hilbert space. The role of quantum characteristics concurs with the role of classical characteristics in solving the classical Liouville equation. Quantum characteristics are physically distinct from trajectories of the de Broglie - Bohm theory \cite{DBB} and trajectories appearing in the semiclassically concentrated states \cite{BAGR1,BAGR2}. 

In Sect. III, we derive the principle of stationary action for quantum Hamilton's equations, extend to quantum systems the Liouville theorem on conservation of the phase-space volume by the classical phase flow, and propose the quantum counterpart of the Poincar\'e theorem on conservation of $2p$ forms. 

Sect. IV is devoted to the semiclassical expansion of a star-function around an ordinary dot-function. Specific features of this technique applied to star-functions of quantum characteristics are discussed in Sect. V. We expand quantum characteristics in a power series in $\hbar$ and derive a system of coupled ODE for components of quantum characteristics of the $\hbar$ expansion and for the associated Jacobi fields. We construct in particular the Green function for the lowest order quantum correction to classical phase-space trajectory in terms of the Jacobi fields.

The lowest order quantum corrections for the classical Kepler periodic orbits are constructed in Sect. VI.

Numerical methods of solution of many-body scattering problems using the $\hbar$ expansion and calculation of average values of physical observables in the course of quantum evolution are discussed in Sect. VII. In Sect. VIII, we summarize results.

\section{Weyl's association rule and the star-product}
\setcounter{equation}{0}

Classical systems with $n$ degrees of freedom are described within the 
Hamiltonian framework by $2n$ canonical coordinates and momenta 
$\xi^{i}=(q^{1},...,q^{n},p_{1},...,p_{n})$ which satisfy the Poisson 
bracket relations 
\begin{equation}
\{\xi^{k},\xi^{l} \} = - I^{kl},
\label{POIS}
\end{equation}
with the matrix $I^{kl}$
\begin{equation}
\left\| I \right\| =\left\| 
\begin{array}{ll}
0 & -E_{n} \\ 
E_{n} & 0
\end{array}
\right\|
\label{MATR}
\end{equation}
where $E_{n}$ is the $n\times n$ identity matrix. In quantum mechanics, 
canonical variables $\xi^{i}$ are associated to operators of canonical 
coordinates and momenta 
$\mathfrak{x}^{i} = (\mathfrak{q}^{1},...,\mathfrak{q}^{n},\mathfrak{p}_{1},...,\mathfrak{p}_{n})
\in Op(L^{2}(\mathbb{R}^{n}))$ 
acting in the Hilbert space, which obey the commutation relations
\begin{equation}
[ \mathfrak{x}^{k},\mathfrak{x}^{l} ] = -i\hbar I^{kl}.
\label{COMM} 
\end{equation}
The Weyl's association rule extends the correspondence 
$\xi^{i} \leftrightarrow \mathfrak{x}^{i}$ to arbitrary phase-space 
functions and operators.

The set of operators $\mathfrak{f}$ acting in the Hilbert space is closed under the multiplication of operators by $c$-numbers and summation of operators. Such a set constitutes the vector space $V$. The elements of its basis can be labelled by canonical variables $\xi^{i}$. The commonly used Weyl's basis looks like
\begin{equation}
\mathfrak{B}(\xi )= (2\pi \hbar )^{n}\delta^{2n}(\xi - \mathfrak{x}) = \int \frac{d^{2n}\eta }{(2\pi \hbar )^{n}}
\exp (-\frac{i}{\hbar }\eta _{k}(\xi - \mathfrak{x})^{k}).  \label{P}
\end{equation}

The Weyl's association rule for a function $f(\xi )$ and an operator 
$\mathfrak{f}$ has the form \cite{STRA}
\begin{eqnarray}
f(\xi ) &=&Tr[\mathfrak{B}(\xi )\mathfrak{f}],  \label{S} \\
\mathfrak{f} &=&\int \frac{d^{2n}\xi }{(2\pi \hbar )^{n}}f(\xi )\mathfrak{B}(\xi ).
\label{INV}
\end{eqnarray}
In particular, $\xi ^{i}=Tr[\mathfrak{B}(\xi )\mathfrak{x}^{i}]$. The function $f(\xi)$ can be treated as the coordinate of $\mathfrak{f}$ in the basis $\mathfrak{B}(\xi )$, Eq.(\ref{S}) as the scalar product of $\mathfrak{B}(\xi )$ and $\mathfrak{f}$. 

Alternative operator bases and their relations are discussed in Refs. \cite{MEHTA,BALAZ}. One can make, in particular, operator transformations on $\mathfrak{B}(\xi )$ and $c$-number transformations on $\xi^{i}$. 

The Weyl-symmetrized functions of operators of canonical variables have the representation \cite{KAMA91}
\begin{equation}
\mathfrak{f} = f(\frac{ \mathfrak{q}_{(1)}^i + \mathfrak{q}_{(3)}^i}{2}, \mathfrak{p}_{(2)}^i),
\label{WEYL3}
\end{equation}
where the subscripts indicate the order in which the operators act on the right.

The set of operators is closed under the multiplication of operators. The vector space $V$ is endowed thereby with an associative algebra structure. Given two functions 
$f(\xi ) = Tr[\mathfrak{B}(\xi )\mathfrak{f}]$ and $g(\xi ) = Tr[\mathfrak{B}(\xi )\mathfrak{g}]$,
one can construct a third function 
\begin{equation}
f(\xi )\star g(\xi )=Tr[\mathfrak{B}(\xi )\mathfrak{fg}]  \label{GR}
\end{equation}
called star-product \cite{GROE}. It is given explicitly by
\begin{equation}
f(\xi )\star g(\xi )=f(\xi )\exp (\frac{i\hbar }{2}\mathcal{P})g(\xi ),
\label{EG}
\end{equation}
where 
\begin{equation}
\mathcal{P}\mathcal{=}-{I}^{kl}\overleftarrow{\frac{%
\partial }{\partial \xi ^{k}}}\overrightarrow{\frac{\partial }{\partial \xi
^{l}}} 
\label{POISOPER}
\end{equation}
is the Poisson operator. 

The star-product splits into symmetric and skew-symmetric parts 
\begin{equation}
f\star g=f\circ g+\frac{i\hbar}{2}  f\wedge g.
\label{STAR}
\end{equation}
The skew-symmetric part $f \wedge g$ is known under the name of Moyal bracket. 

The Weyl's symbol of a symmetrized product $\mathfrak{x}^{i_{1}}\mathfrak{x}^{i_{2}}...\mathfrak{x}%
^{i_{s}}$ simplifies to a pointwise product of canonical variables 
\begin{eqnarray}
Tr[\mathfrak{B}(\xi )\mathfrak{x}^{(i_{1}}\mathfrak{x}^{i_{2}}...\mathfrak{x}^{i_{s})}] = \xi ^{i_{1}}...\xi ^{i_{s}}, 
\label{COOR}
\end{eqnarray}
which is explicitly symmetric with respect to permutations of the indices. 
The symmetrized product of $2n$ Hermitian operators $\mathfrak{u}^{i}$ is associated to the 
symmetrized star-product of real phase-space functions $u^{i}(\xi )=Tr[\mathfrak{B}(\xi )\mathfrak{u}^{i}]$: 
\begin{eqnarray}
Tr[\mathfrak{B}(\xi )\mathfrak{u}^{(i_{2}}\mathfrak{u}^{i_{2}}...\mathfrak{u}^{i_{s})}]
=u^{(i_{1}}(\xi )\circ u^{i_{2}}(\xi )\circ ...\circ u^{i_{s})}(\xi ).
\label{uuuu}
\end{eqnarray}
The $\circ$-product is not associative. The order in which it acts in 
Eq.(\ref{uuuu}) is, however, not important since the indices are symmetrized.

The Weyl's association rule can be reformulated in terms of the Taylor expansion. Consider the Taylor expansion of a function $f(\xi )$: 
\begin{equation}
f(\xi )=\sum_{s=0}^{\infty }\frac{1}{s!}\frac{\partial ^{s}f(0)}{\partial
\xi ^{i_{1}}...\partial \xi ^{i_{s}}}\xi ^{i_{1}}...\xi ^{i_{s}}.
\label{TAYL}
\end{equation}
For any function $f(\xi )$ one may associate an operator $\mathfrak{f}_{T}$: 
\begin{equation}
\mathfrak{f}_{T}=\sum_{s=0}^{\infty }\frac{1}{s!}\frac{\partial ^{s}f(0)}{%
\partial \xi ^{i_{1}}...\partial \xi ^{i_{s}}}\mathfrak{x}^{i_{1}}...\mathfrak{x}%
^{i_{s}}.
\label{TAYLOR}
\end{equation}
The simple calculation 
\begin{eqnarray}
f_{T}(\xi ) = Tr[\mathfrak{B}(\xi )\mathfrak{f}_{T}] 
&=&\sum_{s=0}^{\infty }\frac{1}{s!}\frac{\partial ^{s}f(0)}{\partial \xi
^{i_{1}}...\partial \xi ^{i_{s}}}\xi ^{i_{1}}\star ...\star \xi ^{i_{s}} \nonumber \\
&=&\sum_{s=0}^{\infty }\frac{1}{s!}\frac{\partial ^{s}f(0)}{\partial \xi
^{i_{1}}...\partial \xi ^{i_{s}}}\xi ^{i_{1}}...\xi ^{i_{s}}
\label{FINI}
\end{eqnarray}
shows that $f_{T}(\xi )=f(\xi )$ and $\mathfrak{f}_{T}=%
\mathfrak{f}.$ The Taylor expansion over the symmetrized products of
operators of canonical coordinates and momenta gives the association 
rule equivalent to Eqs.(\ref{S}) and (\ref{INV}). 
So, one can write $\mathfrak{f} = f(\mathfrak{x})$. 

The average values of a physical observable described by an operator 
$\mathfrak{f}$ is calculated as 
trace of the operator product $\mathfrak{fr}$ where $\mathfrak{r}=%
\mathfrak{r}^{+}$ is the density matrix, $Tr[\mathfrak{r}]=1$ or, equivalently, by
averaging the phase-space function $f(\xi )$ over the Wigner function 
\begin{equation}
W(\xi )=Tr[\mathfrak{B}(\xi )\mathfrak{r}].  \label{WIGN}
\end{equation}
The Wigner functions is normalized to unity 
\begin{equation}
\int \frac{d^{2n}\xi }{(2\pi \hbar )^{n}}W(\xi )=1.
\label{WFUN}
\end{equation}
If $\mathfrak{f} \leftrightarrow f(\xi)$ and $\mathfrak{r} \leftrightarrow W(\xi)$, then 
\begin{equation}
Tr[\mathfrak{fr}]=\int \frac{d^{2n}\xi }{(2\pi \hbar)^{n}}f(\xi )\star W(\xi ) = \int \frac{d^{2n}\xi }{(2\pi \hbar)^{n}}f(\xi )W(\xi ).
\label{TR}
\end{equation}
The star-product can be replaced with the pointwise product \cite{STRA,BAYEN,FAIRLI}.

Not every normalized function in phase space can be interpreted as the Wigner function. The eigenvalues of density matrices are positive, so given $%
\mathfrak{r}$, one can find a Hermitian matrix $\mathfrak{r}%
_{1/2}$ such that $\mathfrak{r}=\mathfrak{r}_{1/2}\mathfrak{r}_{1/2}$.
For any $W(\xi )$ there exists therefore $W_{1/2}(\xi )$ such that 
$W(\xi )=W_{1/2}(\xi )\star W_{1/2}(\xi )$.

For a pure state  $\mathfrak{r} = |\psi><\psi|$, the Wigner function becomes 
$W(\xi ) = <\psi|\mathfrak{B}(\xi)|\psi>$. Its value is restricted by $-2^n \leq W(\xi ) \leq 2^{n}$ provided $|\psi>$ has a finite norm \cite{BACKE}.

\section{Quantum characteristics}
\setcounter{equation}{0}       

This section is devoted to studying the transformation properties of the Weyl's symbols of operators under the action of unitary evolution operators. One-parameter set of unitary transformations applied to the Heisenberg operators of canonical variables generates, under the Weyl's association rule, phase-space trajectories. The knowledge of these trajectories is equivalent to the knowledge of quantum dynamics. In particular, time-dependent symbols of operators are functionals of such trajectories. In this sense, the phase-space trajectories play the role of characteristics. The interest to the quantum characteristics is connected with the special role of trajectories in the transport models.

In Euclidean space, we can construct basis $\{\textbf{e}_{i}\}$ and study two kind of transformations: Passive ones where $\textbf{e}_{i}$ are transformed and other vectors do not transform, and active ones where $\textbf{e}_{i}$ are not transformed and other vectors are transformed. These two views are equivalent. In what follows, active transformations are considered, the operator basis $\mathfrak{B}(\xi)$ remains fixed. This corresponds to the Heisenberg picture where evolution applies to operators associated to physical observables. 

\subsection{Unitary transformations under the Weyl's association rule}

Consider unitary transformation of an operator 
$\mathfrak{f}\rightarrow \acute{\mathfrak{f}}=\mathfrak{U^{+}fU}$ where 
$\mathfrak{U}^{+}\mathfrak{U}=\mathfrak{U}\mathfrak{U}^{+}=1.$ The operators 
of canonical variables are transformed as
\[
\mathfrak{x}^{i}\rightarrow \acute{\mathfrak{x}}^{i}=\mathfrak{U^{+}\mathfrak{x}^{i}U}, 
\]
whereas
their symbols as $\xi^{i} \rightarrow \acute{\xi}^{ i}=Tr[\mathfrak{B}(\xi )\mathfrak{U^{+}\mathfrak{x}^{i}U}]$. Define
\begin{equation}
u^{i}(\xi )=Tr[\mathfrak{B}(\xi )\mathfrak{U^{+}\mathfrak{x}^{i}U}].  \label{U}
\end{equation}
The associated transformation of function $f(\xi )$ has the form
\begin{eqnarray}
f(\xi ) \rightarrow \acute{f}(\xi ) &=&Tr[\mathfrak{B}(\xi )\acute{\mathfrak{f}}]=Tr[\mathfrak{B}(\xi )%
\mathfrak{U^{+}fU}]  \nonumber \\
&=&\sum_{s=0}^{\infty }\frac{1}{s!}\frac{\partial ^{s}f(0)}{\partial \xi
^{i_{1}}...\partial \xi ^{i_{s}}}Tr[\mathfrak{B}(\xi )\mathfrak{U^{+}\mathfrak{x}^{i_{1}}...%
\mathfrak{x}^{i_{s}}U}]  \nonumber \\
&=&\sum_{s=0}^{\infty }\frac{1}{s!}\frac{\partial ^{s}f(0)}{\partial \xi
^{i_{1}}...\partial \xi ^{i_{s}}}Tr[\mathfrak{B}(\xi )\acute{\mathfrak{x}}^{i_{1}} ... \acute{\mathfrak{x}}^{i_{s}}]  \nonumber \\
&=&\sum_{s=0}^{\infty }\frac{1}{s!}\frac{\partial ^{s}f(0)}{\partial \xi
^{i_{1}}...\partial \xi ^{i_{s}}}u^{i_{1}}(\xi )\star ...\star u^{i_{s}}(\xi
)  \nonumber \\
&\equiv &f(\star u(\xi )).  \label{TF}
\end{eqnarray}
The $\star$-product can be replaced with the $\circ$-product. The $\circ$-product is not associative. The order in which it acts is, however, not important due to symmetrization over the indices. As a consequence, semiclassical expansion of $f(\star u(\xi )) $ around $f(u(\xi )) $ involves even powers of $\hbar$ only. In general, $f(\star u(\xi)) = f(\circ u(\xi)) \neq f(u(\xi))$, while $f(\star u(\xi)) = f(u(\xi))$ provided $u^{i}(\xi)$ is a linear function in $\xi^{i}$.

The antisymmetrized products $\mathfrak{x}^{[i_{1}}...\mathfrak{x}^{i_{2s}]}$
of even numbers of operators of canonical coordinates and momenta represent 
$c$-numbers. They are left invariant by unitary transformations: 
\begin{equation}
\mathfrak{U}^{+}\mathfrak{x}^{[i_{1}}...\mathfrak{x}^{i_{2s}]}\mathfrak{U}
                             =\mathfrak{x}^{[i_{1}}...\mathfrak{x}^{i_{2s}]}.  \label{QP}
\end{equation}
In phase space, equations (\ref{QP}) look like 
\begin{eqnarray}
u^{[i_{1}}(\xi )\star ...\star u^{i_{2s}]}(\xi ) &=& \xi ^{[i_{1}}\star
...\star \xi ^{i_{2s}]} \nonumber \\ 
&=& \left( \frac{- i\hbar}{2} \right)^{s} \frac{1}{(2s)!}\sum_{\sigma}
(-)^{\sigma}{I}^{i_{1}i_{2}}...{I}^{i_{2s - 1}i_{2s}}  
\label{QPPHASE}
\end{eqnarray}
where summation runs over permutations of indices $(i_{1}i_{2}...i_{2s})$ and $(-)^{\sigma} = \pm 1$ 
depending as the ordered set $(i_{1}i_{2}...i_{2s})$ constitutes even or odd permutation of $(1,2,...,2s)$. In particular, 
\begin{equation}
u^{i}(\xi )\wedge u^{j}(\xi )=\xi ^{i}\wedge \xi ^{j}=- {I}^{ij}. 
\label{AREA}
\end{equation}

One may associate to real functions $u^{i}(\xi )$ Hermitian operators 
$\acute{\mathfrak{x}}^{i} = u^{i}(\mathfrak{x})$. If functions $u^{i}(\xi )$ obey Eqs.(\ref{AREA}), operators $\acute{\mathfrak{x}}^{i}$ obey the commutation rules for operators of canonical coordinates and momenta
\begin{equation}
[u^{i}(\mathfrak{x}),u^{j}(\mathfrak{x})] = [\mathfrak{x}^{i},\mathfrak{x}^{j}] = -i\hbar I^{ij}.
\label{CRUL}
\end{equation}
There exists therefore a unitary operator $\mathfrak{U}$ which relates 
$\mathfrak{x}^{i}$ and $\acute{\mathfrak{x}}^{i}$. 
\footnote{
Strictly speaking, this is valid for those $\mathfrak{U}$ only 
which can continuously be shrunk to a unit (cf. \cite{BLEAF}). This is 
the case we are interested in, as $\mathfrak{U}$ is the evolution operator (\ref{UMAT}).}
Applying Eq.(\ref{TF}) to the product $\mathfrak{fg}$ of two operators, 
we obtain a function $f(\zeta) \star g(\zeta)|_{\zeta = \star u(\xi,\tau)}$ 
associated to the operator
$\mathfrak{U}^{+}(\mathfrak{fg})\mathfrak{U}$ and function $f(\star u(\xi )) \star g(\star u(\xi ))$ associated to the operator $(\mathfrak{U}^{+}\mathfrak{f}\mathfrak{U})(\mathfrak{U}^{+}\mathfrak{g}\mathfrak{U})$. These two operators coincide, so their symbols coincide also:
\begin{equation}
f(\zeta) \star g(\zeta)|_{\zeta = \star u(\xi)} = f(\star u(\xi )) \star g(\star u(\xi )).  
\label{BRINVA}
\end{equation}
The star-products in the left- and right-hand sides act on $\zeta$ and $\xi$, respectively. 

Equation (\ref{BRINVA}) shows that one can compute the star-product in the initial coordinate system and change the variables $\xi \rightarrow \zeta = \star u(\xi)$, or equivalently, change the variables $\xi \rightarrow \zeta = \star u(\xi)$ and compute the star-product. Equation (\ref{BRINVA}) applies separately to the symmetric and anisymmetric parts of the star-product.

Equation (\ref{BRINVA}) makes it possible to calculate the star-product in new unitary equivalent coordinate systems. The functional form of equations constructed with the use of the summation and the star-multiplication operations remains unchanged in all unitary equivalent coordinate systems. The star-product is not invariant under canonical transformations in general \cite{KRFA}.

\subsection{Quantum phase flows generated by one-parameter sets of unitary transformations}

A one-parameter set of unitary transformations in the Hilbert space is parameterized by
\begin{equation}
\mathfrak{U}=\exp (-\frac{i}{\hbar} \mathfrak{H} \tau), 
\label{UMAT}
\end{equation}
with $\mathfrak{H} =\mathfrak{H}^{+}$ being the Hamiltonian. The functions $u^{i}(\xi )$ defined by (\ref{U}) acquire a dependence on the parameter $\tau $, so one can write $u^{i}(\xi ,\tau )$. They specify quantum phase flow which represents quantum analogue of the classical phase flow \cite{ARNO}. In virtue of Eq.(\ref{TF}), the evolution of symbols of Heisenberg operators is entirely determined by $u^{i}(\xi ,\tau )$.

We keep the conventional term 'canonical transformations' for functions preserving the Poisson bracket. Phase-space transformations which preserve the Moyal bracket (\ref{AREA}) are referred to as 'unitary transformations'. This is consistent with the fact of discussing the continuous unitary operators (\ref{UMAT}). Continuous groups of unitary transformations represent the non-trivial quantum deformation of continuous groups of canonical transformations.
As we distinguish between the Poisson and Moyal brackets, we have to distinguish between the canonical and unitary transformations. 

The relationship between the canonical transformations, which are not elements of continuous groups, and transformations in the Hilbert space appears to be more involved \cite{BLEAF,ANDER}.

The energy conservation in the course of evolution along quantum characteristics implies 
\begin{equation}
H(\xi )=H(\star u(\xi ,\tau ))  \label{EC}
\end{equation}
where 
\begin{equation}
H(\xi )=Tr[\mathfrak{B}(\xi )\mathfrak{H}]
\label{DEFI}
\end{equation}
is the Hamiltonian function. Quantum characteristics can be found from the Hamilton's equations in quantum form (see e.g. \cite{KRFA})
\begin{equation}
\frac{\partial }{\partial \tau }u^{i}(\xi ,\tau ) = \{\zeta ^{i},H(\zeta )\}|_{\zeta =\star u(\xi ,\tau )}  \label{QF}
\end{equation}
with initial conditions 
\begin{equation}
u^{i}(\xi ,0 ) = \xi^{i}. 
\label{INICON}
\end{equation}
The substitution $\zeta =\star u(\xi ,\tau )$ to $F^{i}(\zeta) = \{\zeta ^{i},H(\zeta )\}$ leads to a deformation of the classical phase flow. If the $\star$-symbol would be missing, the quantum effects could be missing also.

The phase-space velocity $\partial u^{i}(\xi ,\tau )/\partial \tau $ depends on $u^{i}(\xi ,\tau )$ like in classical mechanics and on derivatives of $u^{i}(\xi ,\tau )$ with respect to $\xi^{i}$, as a specific manifestation of quantum non-locality.

Functions which stand for physical observables evolve according to equation
\begin{equation}
f(\xi ,\tau ) =Tr[\mathfrak{B}(\xi )\mathfrak{U}^{+}\mathfrak{f}\mathfrak{U}]=f(\star u(\xi ,\tau ),0),  \label{HF}
\end{equation}
whereas the Wigner function does not evolve $W(\xi ,\tau ) = W(\xi ,0 )$. Functions satisfy equation
\begin{equation}
\frac{\partial }{\partial \tau }f(\xi ,\tau )=f(\xi ,\tau )\wedge H(\xi ) = f(\zeta ,0)\wedge H(\zeta )|_{\zeta =\star u(\xi ,\tau )} 
\label{EVOL}
\end{equation}
which is the Weyl's transform of the quantum-mechanical equation of motion for Heisenberg operators.

The formal solutions to Eqs.(\ref{QF}) and (\ref{EVOL}) in the form of series expansions in $\tau $ are given by 
\begin{eqnarray}
u^{i}(\xi ,\tau ) &=&\sum_{s=0}^{\infty }\frac{\tau ^{s}}{s!}\underbrace{(...((}%
_{s}\xi^{i} \wedge H(\xi ))\wedge H(\xi ))\wedge ...H(\xi )),  \label{TEU} \\
f(\star u(\xi ,\tau )) 
&=&\sum_{s=0}^{\infty }\frac{\tau ^{s}}{s!}\underbrace{(...((}_{s}f(\xi
)\wedge H(\xi ))\wedge H(\xi ))\wedge ...H(\xi )).  \label{TEF}
\end{eqnarray}

The transformation of canonical variables to order $O(\tau)$ has the canonical form, since $\xi^{i} \wedge H(\xi ) = \{\xi^{i} , H(\xi )\}$. The second order in $\tau$ gives deviations from the canonical transformation. The infinitesimal transformations generate canonical or unitary finite transformations 
according as the composition law is $u^{i}(u(\xi,\tau_{1}),\tau_{2}) = u^{i}(\xi,\tau_{1} + \tau_{2})$ or $u^{i}(\star u(\xi,\tau_{1}),\tau_{2}) = u^{i}(\xi,\tau_{1} + \tau_{2})$, respectively. 
The infinitesimal transformations of functions which are not linear in $\xi^{i}$ are not canonical and cannot result from a $c$-number transformation of the coordinate system.

If an operator $\mathfrak{A}$ commutes with $\mathfrak{H}$, its symbols is conserved in the sense of $A(\xi) = A(\star u(\xi,\tau))$, similarly to Eq.(\ref{EC}). In the Schr\"odinger picture, the Wigner function of a stationary state commutes with $\mathfrak{H}$ and obeys $W_{S}(\xi) = W_{S}(\star u(\xi,-\tau))$. For a harmonic oscillator, $u^{i}(\xi,\tau)$ depend linearly on $\xi^{i}$ and coincide with the classical trajectories \cite{GROE,BARL,LESCH}. In such a case, the $\star$-symbol can be dropped to give $W_{S}(\xi) = W_{S}(u(\xi,-\tau))$. 

\subsection{Green function in phase space in terms of characteristics}

We wish to establish a connection of the phase-space Green function \cite{BLEAF1,MARIN} with quantum characteristics and to derive Eqs.(\ref{QF}) using the Green function method.

Combining Eq.(\ref{S}) and Eq.(\ref{INV}), one gets
\begin{equation}
\mathfrak{f} = \int \frac{d^{2n}\xi}{(2\pi\hbar)^{n}} \mathfrak{B}(\xi)Tr[\mathfrak{B}(\xi)\mathfrak{f}].
\label{IV6}
\end{equation}
In Eq.(\ref{HF}), we substitute on place of $\mathfrak{f}$ the right-hand side 
of Eq.(\ref{IV6}) and obtain
\begin{equation}
f(\xi,\tau) = \int \frac{d^{2n}\zeta}{(2\pi\hbar)^{n}} D(\xi,\zeta,\tau)f(\zeta,0)
\label{EVVV1}
\end{equation}
where $f(\xi,0) \equiv f(\xi)$ and
\begin{equation}
D(\xi,\zeta,\tau) = D(\zeta,\xi,-\tau) = Tr[\mathfrak{B}(\xi)\mathfrak{U}^{+}\mathfrak{B}(\zeta)\mathfrak{U}] 
= Tr[\mathfrak{B}(\zeta)\mathfrak{U}\mathfrak{B}(\xi)\mathfrak{U}^{+}],
\label{EVVV2}
\end{equation}
with $D(\xi,\zeta,\tau)$ being the Green function. The use of 
Eqs.(\ref{IV6}) and (\ref{EVVV2}) for $\mathfrak{U} = \mathfrak{U}_{2}\mathfrak{U}_{1}$ gives
\begin{equation}
\int \frac{d^{2n}\zeta}{(2\pi \hbar)^{n}}D(\xi_{1},\zeta,\tau_{1})D(\zeta,\xi_{2},\tau_{2})
 = D(\xi_{1},\xi_{2},\tau_{1} + \tau_{2}).
\label{EVVV3}
\end{equation}
The Green function can be treated as the Weyl's symbol of $\mathfrak{U}^{+}\mathfrak{B}(\zeta)\mathfrak{U}$ 
in the basis 
$\mathfrak{B}(\xi)$ or the Weyl's symbol of $\mathfrak{U}\mathfrak{B}(\xi)\mathfrak{U}^{+}$ in the basis 
$\mathfrak{B}(\zeta)$. The orthogonality condition
\begin{equation}
Tr[\mathfrak{B}(\xi)\mathfrak{B}(\zeta)] = (2\pi \hbar)^{n}\delta^{2n}(\xi - \zeta)
\label{ORTH}
\end{equation}
follows from Eqs.(\ref{S}) and (\ref{INV}). Using Eq.(\ref{TF}), we obtain
\begin{equation}
D(\xi,\zeta,\tau) = (2\pi \hbar)^{n}\delta^{2n}(\star u(\xi,\tau) - \zeta) = (2\pi \hbar)^{n}\delta^{2n}(\xi - \star u(\zeta,-\tau)).
\label{EVVV41}
\end{equation}

The evolution equation for the Green function can be found by calculating the time derivative of Eq.(\ref{EVVV2}) and applying the Weyl's transform to the commutators of $\mathfrak{H}$ and $\mathfrak{U}^{+}\mathfrak{B}(\zeta)\mathfrak{U}$ or $\mathfrak{U}\mathfrak{B}(\xi)\mathfrak{U}^{+}$:
\begin{equation}
\frac{\partial}{\partial \tau} D(\xi,\zeta,\tau)  =   D(\xi,\zeta,\tau) \wedge H(\xi) 
                                                         = - D(\xi,\zeta,\tau) \wedge H(\zeta). \label{EV1} \\
\end{equation}

One can replace in Eq.(\ref{EVVV1}) $f(\zeta,0)$ by $\zeta^{i}$ and $f(\xi,\tau)$ by $u^{i}(\xi,\tau)$, take the time derivative and apply Eq.(\ref{EV1}). We get
\begin{eqnarray}
\frac{\partial }{\partial \tau }u^{i}(\xi ,\tau ) = - \int \frac{d^{2n}\zeta}{(2\pi \hbar)^{2n}} \left( D(\xi,\zeta,\tau) \wedge H(\zeta) \right) \zeta^{i} &=& 
- \int \frac{d^{2n}\zeta}{(2\pi \hbar)^{2n}} \left( D(\xi,\zeta,\tau) \wedge H(\zeta) \right) \star \zeta^{i} \nonumber \\
= \int \frac{d^{2n}\zeta}{(2\pi \hbar)^{2n}} D(\xi,\zeta,\tau)  \star \left(\zeta^{i} \wedge H(\zeta) \right) &=& 
\int \frac{d^{2n}\zeta}{(2\pi \hbar)^{2n}} D(\xi,\zeta,\tau) \left(\zeta^{i} \wedge H(\zeta) \right).
\label{NULL}
\end{eqnarray}
Using the identity $\zeta^{i} \wedge H(\zeta) = \{\zeta^{i} , H(\zeta) \}$ and the explicit form of the Green function (\ref{EVVV41}) we arrive at Eqs.(\ref{QF}).


\subsection{Principle of stationary action for quantum Hamilton's equations}


The quantum Hamilton's equations can be derived from the principle of the stationary action. The action $S$ is the integral 
\begin{equation}
S = \int_{\tau_{1}}^{\tau_{2}} d\tau \int \frac{d^{2n}\xi}{(2\pi \hbar)^{n}%
}  \left( - \frac{1}{2} I_{ij} u^{i}(\xi,\tau) \star \dot{u}%
^{j}(\xi,\tau) - H(\star u(\xi,\tau)) \right). 
\label{LABE1}
\end{equation}
We compare trajectories $u(\xi,\tau)$ coming out $\xi$ at $\tau_{1}$. The action is stationary provided $2n$ functions $u^{i}(\xi,\tau)$ obey equations (\ref{QF}).

The stationary action principle can be reformulated in more familiar terms for Heisenberg operators of the canonical coordinates and momenta: 
\begin{eqnarray}
S = \int_{\tau_{1}}^{\tau_{2}} d\tau \;Tr \left[ - \frac{1}{2} I_{ij} %
\mathfrak{x}^{i} \dot{\mathfrak{x}}^{j} - H(\mathfrak{x}) \right].
\label{LABE2}
\end{eqnarray}
We recall that $H(\mathfrak{x}) = \mathfrak{H}$ is the Hamiltonian. Using 
\begin{equation}
[\mathfrak{x}^{i},H(\mathfrak{x})] = i\hbar \{\xi^{i},H(\xi)\}|_{\xi = %
\mathfrak{x}},  \label{1111}
\end{equation}
one arrives at the Heisenberg equations of motion 
\begin{equation}
i\hbar \frac{\partial}{\partial \tau} \mathfrak{x}^{i} = [\mathfrak{x}^{i},H(%
\mathfrak{x})]  \label{2222}
\end{equation}
and, further, at Eq.(\ref{QF}) upon the Weyl's transform.


\subsection{Quantum counterpart of the Liouville theorem}


The Liouville theorem states conservation of the phase-space volume by the classical phase flow. The Poincar{\'e} theorem of Hamiltonian dynamics suggests conservation of the $2p$-forms under the canonical transformations (see
e.g. \cite{ARNO}). We describe quantum analogues of these remarkable theorems.

\vspace{8mm} 
\begin{figure}[!htb]
\begin{center}
\includegraphics[angle=270,width=5.5 cm]{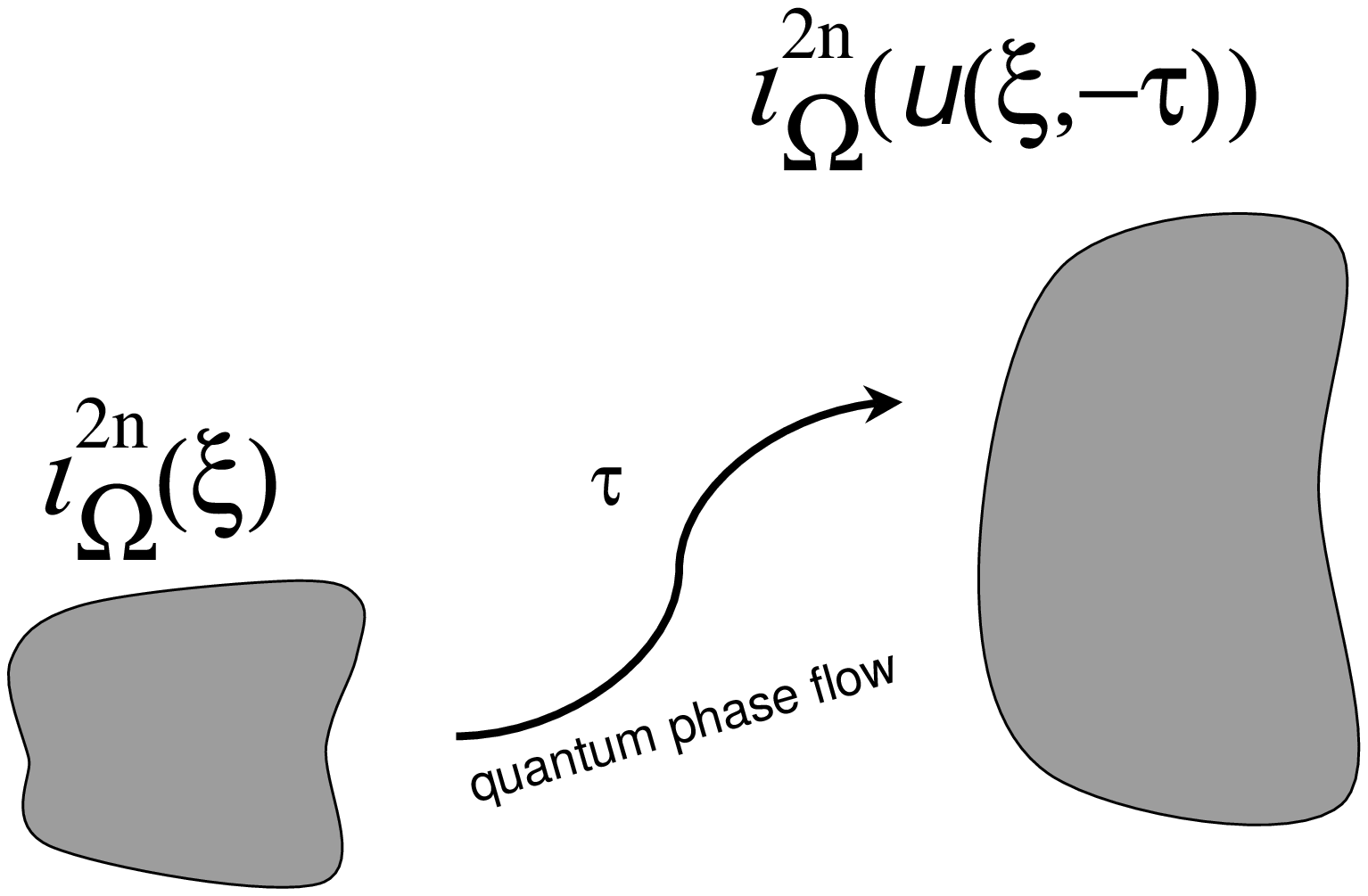}
\end{center}
\caption{Quantum phase flow applied classically to region $\Omega$ 
in phase space. The phase-space volume is not conserved with time. 
The region $\Omega$ at $\tau = 0$ is specified by its index function 
$i_{\Omega}^{2n}(\xi)$. The region $\Omega$ at $\tau \neq 0$ is 
specified by the evolved index function $i_{\Omega}^{2n}(u(\xi,-\tau))$.
}
\label{fig9}
\end{figure}

Let $\Omega$ be a region in phase space. Its index function is defined by 
\begin{equation}
i_{\Omega}^{2n}(\xi)=\left\{ 
\begin{array}{l}
1,\;\;\; \xi \in \Omega, \\ 
0,\;\;\; \xi \notin \Omega.
\end{array}
\right.  \label{SIGN}
\end{equation}

The function $i_{\Omega}^{2n}(\xi)$ can be associated to a region $\Omega$
smaller than it is allowed by the uncertainty principle. It does not fulfill
the condition of divisibility satisfied by the Wigner functions: For every $%
W(\xi)$ one can find $W_{1/2}(\xi)$ such that $W(\xi) = W_{1/2}(\xi)\star
W_{1/2}(\xi)$. The index function does not belong to the set 
of continuous functions, $C^{\infty}(T_{*}\mathbb{R}^{n})$, whose derivatives all 
are continuous. There are no reasons to restrict with the set 
$C^{\infty}(T_{*}\mathbb{R}^{n})$, however, since infinite sequences of continuous 
functions have limits which are not continuous functions in general.
Applying the Weyl's association rule to $i_{\Omega}^{2n}(\xi)$, we obtain 
a Hermitian operator $\mathfrak{i}_{\Omega}^{2n}$.

The phase-space volume occupied by $\Omega$ at $\tau = 0$ is calculated
classically 
\begin{equation}
\Sigma^{2n}(0) = \int \frac{d^{2n}\xi}{(2\pi \hbar)^{n}}i_{\Omega}^{2n}(\xi) = Tr[\mathfrak{i}_{\Omega}^{2n}].  
\label{VOLUME}
\end{equation}

Quantum phase flow preserves the Moyal bracket rather than the Poisson
bracket and does thereby not transform canonically. The evolution of $\Omega$
along quantum characteristics, respectively, does not preserve the
phase-space volume: 
\begin{equation}
\Sigma^{2n}_{c}(\tau) = \int \frac{d^{2n}\xi}{(2\pi \hbar)^{n}}i_{\Omega}^{2n}(u(\xi,-\tau)) 
\neq \Sigma^{2n}(0).  \label{CLASS1}
\end{equation}
The variance appears at the order $O(\hbar^2)$.

\vspace{12mm} 
\begin{figure}[!htb]
\begin{center}
\includegraphics[angle=270,width=6.0 cm]{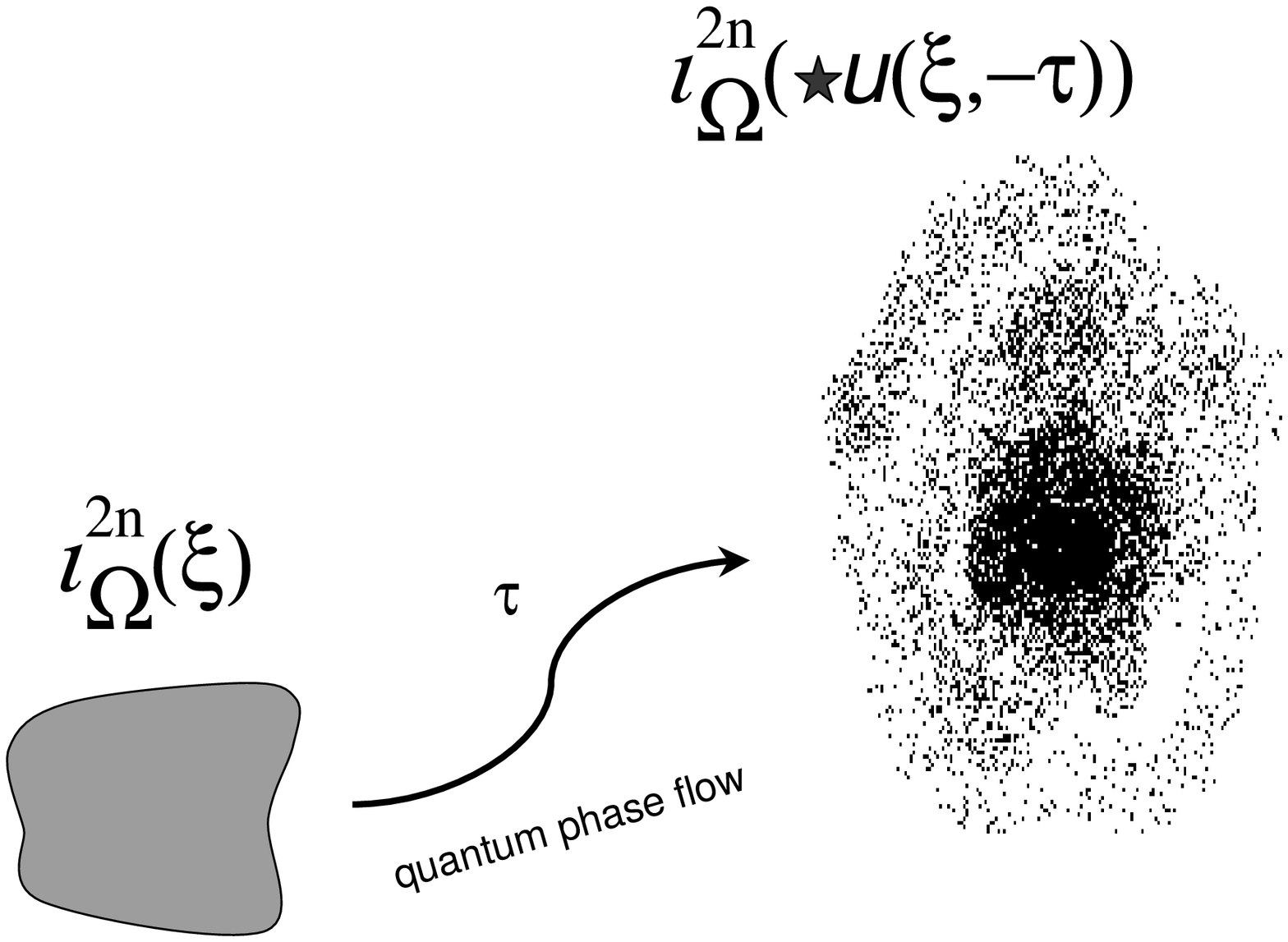}
\end{center}
\caption{A visual demonstration of quantum evolution of desity
distribution associated to index function $i_{\Omega}^{2n}(\xi)$ 
in phase space at $\tau = 0$. 
The boundary of region $\Omega$ diffuses in the course of quantum 
evolution. The integral of the evolved index function 
$i_{\Omega}^{2n}(\star u(\xi,-\tau))$ over phase space is 
conserved. }
\label{fig8}
\end{figure}

One can make a positive statement, however. Implicitly, we treated the
transform classically substituting $i_{\Omega}^{2n}(\xi) \rightarrow
i_{\Omega}^{2n}(u(\xi,-\tau))$. Quantum particles, however, do not move
along quantum characteristics, as must be clear from the composition law 
\cite{KRFA}
\begin{equation}
u(\xi,\tau_{1}+\tau_{2}) = u(\star u(\xi,\tau_{1}),\tau_{2}).
\label{comp}
\end{equation}
The boundary of $\Omega$ does, naturally, not move along
quantum characteristics also. It experiences quantum fluctuations. In the
spirit of quantum mechanics, one can speak of conservation of average
volume only. The quantum evolution is formally expressed through the
star-product transform $\xi \rightarrow \star u(\xi,-\tau)$, so we have to
write 
\begin{equation}
\Sigma^{2n}(\tau) = \int \frac{d^{2n}\xi}{(2\pi \hbar)^{n}}%
i_{\Omega}^{2n}(\star u(\xi,-\tau))
= Tr[\mathfrak{U} \mathfrak{i}_{\Omega}^{2n}\mathfrak{U}^{+}].
\label{CLASS}
\end{equation}
Equation 
\begin{equation}
\frac{\partial }{\partial \tau} Tr[\mathfrak{f}] = 0  \label{QLE}
\end{equation}
is valid for any Heisenberg operator $\mathfrak{f} \in Op(L^{2}(\mathbb{R}^{n}))$
including $\mathfrak{i}_{\Omega}^{2n}$.

Remarkably, it incorporates the phase-space volume conservation within the
framework of the Groenewold-Moyal dynamics: 
\begin{equation}
\Sigma^{2n}(\tau) = \Sigma^{2n}(0).  \label{CONSE}
\end{equation}
In the limit of $\hbar \rightarrow 0$, $u(\xi,\tau) \rightarrow c(\xi,\tau)$
where $c(\xi,\tau)$ are solutions of the classical Hamilton's equations. One has
therefore $\lim_{\hbar \rightarrow 0} \Sigma^{2n}(\tau) = \lim_{\hbar
\rightarrow 0} \Sigma^{2n}_{c}(\tau)$, and we recover the classical
Liouville theorem. Equation (\ref{QLE}) and its consequence Eq.(\ref{CONSE}) 
comprise the quantum-mechanical analogue of the classical
Liouville theorem.


\subsection{Quantum counterpart of the Poincar\'e theorem on conservation of 
$2p$-forms}


The quantum analogue of the Poincar\'e theorem on conservation of the $2p$%
-forms under canonical transformations follows from Eq.(\ref{QLE}) also.

Let $\Omega_{2p}$ be a $2p$ dimensional submanifold in phase space. It can
be specified by constraint equations $\mathcal{G}^{a}(\xi) = 0$ for $a =
1,...,2m$ where $m=n-p$.

The volume element of $\Omega_{2p}$ is given by \cite{FADD} 
\begin{equation}
d\Sigma^{2p}(\xi) = \frac{d^{2n}\xi}{(2\pi \hbar)^{n}} (\det \{\mathcal{G}%
^{a}(\xi),\mathcal{G}^{b}(\xi)\} )^{1/2}
(2\pi \hbar)^{m} \prod_{c=1}^{2m}\delta(\mathcal{G}^{c}(\xi)).
\label{MEASU}
\end{equation}

The index function $i_{\Omega}^{2n}(\xi)$ of $\Omega_{2p}$ can be taken to
be 
\begin{equation}
i_{\Omega}^{2p}(\xi) =  (\det \{%
\mathcal{G}^{a}(\xi),\mathcal{G}^{b}(\xi)\} )^{1/2} 
(2\pi \hbar)^{m} \prod_{c=1}^{2m}\delta(\mathcal{G}%
^{c}(\xi))i_{\Omega}^{\prime 2n}(\xi).  \label{UFFFF}
\end{equation}
where $i_{\Omega}^{\prime 2n}(\xi)$ is an index function in phase space
non-singular across the constraint submanifold. The delta-functions force
$i_{\Omega}^{2p}(\xi)$ to vanish outside
of $\Omega_{2p}$.

The function (\ref{UFFFF}) can be associated to an index operator $%
\mathfrak{i}_{\Omega}^{2p}$ by the Weyl's association rule Eq.(\ref{INV}). The $2p$ dimensional volume $%
\Sigma^{2p}(\tau)$ has the form of Eq.(\ref{CLASS}) with $%
i_{\Omega}^{2n}(\xi)$ replaced by $i_{\Omega}^{2p}(\xi)$.

The quantum phase flow, in virtue of Eq.(\ref{QLE}), preserves the $2p$
volume 
\begin{equation}
\Sigma^{2p}(\tau) = \Sigma^{2p}(0).  \label{POINC}
\end{equation}
For $p=n$, we recover Eq.(\ref{CONSE}).

In classical mechanics, $2p$-forms take values on vectors $%
d\xi_{1},d\xi_{2},...,d\xi_{2p}$ to give oriented volumes of $2p$%
-dimensional parallelepipeds. In the course of evolution, infinitesimal
parallelepipeds remain infinitesimal parallelepipeds.

In quantum mechanics, we can construct infinitesimal parallelepipeds at $%
\tau = 0$ and calculate their volumes also. As distinct from the classical
case, however, quantum evolution does not keep index functions $%
i_{\Omega}^{2p}(\xi,0)$ within the set of index functions. For $\tau \neq 0$%
, $i_{\Omega}^{2p}(\xi,\tau)$ does not correspond to any region in phase
space, although phase-space integral of $i_{\Omega}^{2p}(\xi,\tau)$ is equal
to volume (or area) of the initial parallelepiped at $\tau = 0$. There is
only one instant of time $\tau = 0$ when the formalism shows the clear
geometric sense.

\section{Semiclassical expansion of $f(\star u(\xi,\tau))$ around $f( u(\xi,\tau))$}
\setcounter{equation}{0}

The evolution problem for quantum systems can be split in two parts: First, we look for quantum characteristics which are solutions of Eq.(\ref{QF}) and, secondly, use equation $f(\xi,\tau) = f(\star u(\xi,\tau),0)$ to calculate 
time-dependent symbols of operators. 
The key problem is to have an efficient algorithm to calculate functions of $\zeta = \star u(\xi,\tau)$.
The $\star$-arguments appear in the quantum Hamilton's equations (\ref{QF}) and the evolution equation for functions (\ref{EVOL}).

We give semiclassical expansion of $f(\star u(\xi,\tau))$ around $f(u(\xi,\tau))$. Function $f(\xi)$ can be represented 
as the Fourier transform
\begin{equation}
f(\xi) = \int \frac{d^{2n}\eta}{(2\pi \hbar)^{n}} \exp(\frac{i}{\hbar}\eta_{k}\xi^{k})f(\eta).
\label{FOUR}
\end{equation}
It is sufficient to learn how to calculate $\exp(\star U)$ where 
\[
U = \frac{i}{\hbar}\eta_{k} u^{k}(\xi,\tau).
\]

Using Eq.(\ref{TF}), we obtain \footnote{The MAPLE code for calculation of 
the expansion (\ref{EXPAND}) is available upon request.}
\begin{equation}
\exp( \star U) = \left(1 + \hbar^2 c_2 + \hbar^4 c_4 + O(\hbar^6)\right)\exp(U)
\label{EXPA}
\end{equation}
where 
\begin{eqnarray}
c_2 &=& - \frac{1}{48}(2UUP^2U + 3UP^2U), \label{EXPAND} \\ 
c_4 &=& \frac{1}{23040}( 90(UP^2U)UP^2U + 60(UP^2U)P^2U + 48(UUP^2U)UP^2U
+ 45(UP^2U)(UP^2U) \nonumber \\
&+& 60(UUP^2U)(UP^2U) + 20(UUP^2U)(UUP^2U)  + 30(UUP^2U)P^2U)\nonumber \\
&+& \frac{1}{11520}( 6 UUUUP^4U + 45 UUUP^4U + 30(UUP^4U)_1  + 40(UUP^4U)_2  + 15 UP^4U).
\nonumber 
\end{eqnarray}
Owing to combinatorial factors, $P$ acts like $\mathcal{P}$: In particular, we have $UP^2U = U\mathcal{P}^2U$. 
In the expression $UUP^2U$, $\mathcal{P}^{2}$ acts one time to each $U$ on the left and two times on the right. More sophisticated expression $(UP^2U)P^2U$ is computed by acting $\mathcal{P}^2$ two times on $(UP^2U)$ and two times on $U$. Expressions of the type $ABP^2C$ are computed by acting $\mathcal{P}$ one time on $A$, one time on $B$ and two times on $C$, e.g., $(UP^2U)UP^2U$ stands for $A=(UP^2U)$ and $B=C=U$. The formal description uses notations 
\begin{eqnarray}
A(\xi)_{,i_{1}...i_{s}} &=& \frac{\partial^{s}A(\xi)}{\partial \xi^{i_{1}}...\partial \xi^{i_{s}}}, \label{LIFT1} \\
A(\xi)^{,i_{1}...i_{s}} &=& (-)^{s}I^{i_1 j_1} ...I^{i_s j_s} A(\xi)_{,j_{1}...j_{s}}. \label{LIFT2}
\end{eqnarray}
We have
\begin{eqnarray}
AP^2B &=&A_{,kl}B^{,kl},  \nonumber \\
ABP^2C &=& A_{,k}B_{,l}C^{,kl}, \nonumber \\
AP^4B &=& A_{,ijkl}B^{,ijkl}, 		\nonumber \\
(ABP^4C)_1 &=&A_{,ij}B_{,kl}C^{,ijkl}, \nonumber \\
(ABP^4C)_2 &=&A_{,i}B_{,jkl}C^{,ijkl}, \nonumber \\
ABCP^4D &=&A_{,i}B_{,j}C_{,kl}D^{,ijkl}, \nonumber \\
ABCDP^4E &=& A_{,i}B_{,j}C_{,k}D_{,l}E^{,ijkl}.
\label{DEF6}
\end{eqnarray}

The diagram technique \cite{GRACIA} can be useful for high-order $\hbar$-expansions. Our calculation (\ref{EXPAND}) is in agreement with \cite{GRACIA}.  

The lowest order quantum correction to $f(\star u(\xi,\tau))$ can be found using Eq.(\ref{EXPA}). We replace
$\eta_{i} \rightarrow -i\hbar \partial/\partial u^{i}$ and $U \rightarrow u^{i}(\xi)\partial/\partial u^{i}$ and obtain
\begin{eqnarray}
f(\star u(\xi,\tau)) = f(u(\xi,\tau)) &-& \frac{\hbar^2}{24}u^{i}(\xi,\tau)_{,l}u^{j}(\xi,\tau)_{,m}u^{k}(\xi,\tau)^{,lm}f(u(\xi,\tau))_{,ijk} \nonumber \\
 &-& \frac{\hbar^2}{16}u^{i}(\xi,\tau)_{,kl}u^{j}(\xi,\tau)^{,kl}f(u(\xi,\tau))_{,ij} + O(\hbar^4).
\label{FINA2}
\end{eqnarray} 
The next-order correction can be found using Eqs.(\ref{EXPA}) and (\ref{EXPAND}). The derivatives 
of $f(u(\xi,\tau))$ are calculated with respect to $u^{i}$.

As an application, we consider the semiclassical expansion of the Weyl's symbol of the projection operator $\mathfrak{P}(E) = \delta(E - \mathfrak{H})$. The result is in agreement with Ref. \cite{BALAZ}. The series expansion (\ref{EXPA}) with $U=- H(\xi)/T$ can be used for calculation of the Weyl's symbol of the finite-temperature density matrix.

\section{Semiclassical expansion of quantum characteristics}

In this section, we discuss general properties of the quantum characteristics, specific for the semiclassical expansion.

\subsection{Semiclassical expansion of quantum Hamilton's equations}

The first step in solving evolution problem is to construct quantum characteristics using
Eqs.(\ref{QF}). We make an expansion  in a power series of the Planck's constant: 
\begin{equation}
u^{i}(\xi,\tau) = \sum_{s=0}^{\infty}\hbar^{2s}u^{i}_{s}(\xi,\tau).
\label{ITER1}
\end{equation}
Here, $u^{i}_{0}(\xi,\tau)$ is the classical trajectory which starts at $\tau = 0$ at 
a point $\xi^{i}$. The right-hand side of quantum Hamilton's equations (\ref{QF}), 
$F^{i}(\zeta) = \{ \zeta^{i},H(\zeta)\}$,
is a function of $\zeta = \star u(\xi,\tau)$, so we have to use a power series 
expansion similar to (\ref{FINA2}):
\begin{equation}
F^{i}(\star u(\xi,\tau)) = \sum_{s=0}^{\infty}\hbar^{2s}F^{i}_{s}(u_{0}(\xi,\tau),...,u_{s}(\xi,\tau)).
\label{ITER2}
\end{equation}
Given $ u(\xi,\tau)$, functions $F^{i}_{s}(u_{0}(\xi,\tau),...,u_{s}(\xi,\tau))$ are 
known. $F^{i}_{s}$ depend on derivatives of $u_{r}^{i}(\xi,\tau)$ for $0 \leq r < s$ 
with respect to $\xi^{i}$. In particular,
\begin{eqnarray}
F^{i}_{0}(u_{0}) &=& F^{i}(u_{0}), \nonumber \\
F^{i}_{1}(u_{0},u_{1}) &=&  u^{j}_{1}(\xi,\tau)F^{i}(u_{0})_{,j} - \frac{1}{24}u^{j}_{0}(\xi,\tau)_{,m}u^{k}_{0}(\xi,\tau)_{,n}u^{l}_{0}(\xi,\tau)^{,mn}F^{i}(u_{0})_{,jkl} \nonumber \\
 &-& \frac{1}{16}u^{j}_{0}(\xi,\tau)_{,lm}u^{k}_{0}(\xi,\tau)^{,lm}F^{i}(u_{0})_{,jk},\label{FORCE} 
\end{eqnarray}
and so on. To any fixed order in the Planck's constant, quantum characteristics can be 
found by solving finite-order partial differential equations (PDE)
\begin{eqnarray}
\frac{\partial }{\partial \tau }u^{i}_{s} &=& F^{i}_{s}(u_{0},...,u_{s}), \label{RECU}
\end{eqnarray}
with initial conditions
\begin{eqnarray}
u^{i}_{0}(\xi,0) &=& \xi^{i}, \nonumber \\
u^{i}_{s}(\xi,0) &=& 0, \;\;\; s \geq 1. \label{IRECU}
\end{eqnarray}
Given $u^{i}_{r}(\xi,\tau)$ for $0 \leq r < s$, equations for 
$u^{i}_{s}(\xi,\tau)$ simplify to a system of first-order ordinary 
differential equations (ODE). Such a circumstance allows to approach 
the problem recursively using numerically efficient ODE integrators.

Classical trajectories $u^{i}_{0}(\xi,\tau)$ appear at $O(\hbar^0)$.
To order  $\hbar^2$, the first and second order derivatives of $u^{i}_{0}(\xi,\tau)$ 
with respect to canonical variables $\xi^{i}$ are involved. We have therefore to consider not only propagation of points in phase space, like in classical mechanics, but also propagation of gradients 
\begin{equation}
J^{i...}_{r,.ið_1...ið_t}(\xi,\tau) = \frac{\partial^{t} u^{i}_{r}(\xi,\tau)}{\partial \xi^{i_{1}} ... \partial \xi^{i_{t}}}
\end{equation}
at $0\leq r < s$ which affect high-order quantum corrections $u^{i}_{s}(\xi,\tau)$ and enter semiclassical expansion of $f(\star u(\xi,\tau))$. We call such gradients generalized Jacobi fields (tensors) or simply Jacobi fields, since first-order derivatives, which determine stability of classical trajectories with respect to small perturbations of parameters, 
are known in classical mechanics as the Jacobi fields. We use the symplectic form $I^{ik}$ to shift the indices up and down according to the rule (\ref{LIFT2}), e.g., $x^{i} = -I^{ik}x_{k}$, $x_{i}=-I_{ik}x^{k}$ 
where $I_{kl} = -I^{kl}$, so that $x_{i}y^{i}=-x^{i}y_{i}$. In what follows, we discuss $r=0$ Jacobi fields,
so the lower index $r$ will be suppressed.

\subsection{Semiclassical expansion of energy conservation condition}

The energy conservation in quantum form (\ref{EC}) gives a sequence of conserved quantities
\begin{equation}
H_{s}(u_{0}(\xi,\tau),...,u_{s}(\xi,\tau)) = \lim_{\hbar \rightarrow 0}\frac{1}{(2s)!}\frac{\partial ^{2s}}{\partial \hbar^{2s}}H(\star u(\xi,\tau)).
\label{ECHB}
\end{equation}
For $s \geq 1$ they vanish at $\tau = 0$, so that 
\begin{equation}
H_{s}(u_{0},...,u_{s}) = 0
\label{CONS}
\end{equation}
provided $H(\xi)$ does not depend on $\hbar$. The term $s=0$ in Eq.(\ref{ECHB}) gives the energy conservation law for classical systems (cf. (\ref{EC}))
\begin{equation}
H(\xi)= H(u_{0}(\xi,\tau)).
\end{equation} 

The functions $H_{s}$ and $F_{s}^{i}$ depend linearly on $u^{i}_{s}$ for $s \geq 1$:
$H_{s}(u_{0},...,u_{s}) = u^{i}_{s}H(u_{0})_{,i} + \tilde{H}_{s}(u_{0},...,u_{s-1})$ and 
$F_{s}^{i}(u_{0},...,u_{s}) = -I^{ik}u^{j}_{s}H(u_{0})_{,kj} + \tilde{F}_{s}^{i}(u_{0},...,u_{s-1})$. 
Using equations of motion, one can remove $u^{i}_{s}$ from time derivative of $H_{s}$, so that
\begin{equation}
\frac{\partial }{\partial \tau}H_{s}(u_{0},...,u_{s}) = G_{s}(u_{0},...,u_{s - 1}).
\label{ECHB2}
\end{equation}
In case of $s = 1$, the quantity
\begin{eqnarray}
G_{1}(u_{0}) &=& - \frac{1}{24} \{ J^{i.}_{.l}(\xi,\tau) J^{j.}_{.m}(\xi,\tau) J^{klm}_{...}(\xi,\tau),H(\xi)\}
H(u_{0})_{,ijk} \nonumber \\
 &-& \frac{1}{16} \{ J^{i..}_{.kl}(\xi,\tau) J^{jkl}_{...}(\xi,\tau),H(\xi)\} H(u_{0})_{,ij}
\label{ECHB3}
\end{eqnarray}
remains constant $G_{1}(u_{0}) = 0$ on classical trajectories. It does not depend 
on quantum corrections $u^{i}_{s}(\xi,\tau)$, although involves derivatives 
of $u_{0}^{i}(\xi,\tau)$ with respect to $\xi^{i}$. We have not found for 
$G_{1}(u_{0})$ known counterpart in the classical Hamiltonian theory.
\footnote{Using MAPLE, we verified for $H = (p^2 + q^2)^2$ and a few other 
systems that in the course of evolution $G_{1}(u_{0})$ remains zero at 
least to order $O(\tau^{20})$. For $H = (p^2 + q^2)^2$, we examined  
$H_{s}(u_{0},..,u_{s})$ for $1 \leq s \leq 4$ and found them to be consistent with zero at least to order $O(\tau^{10})$.
First components $s \geq 1$ of quantum characteristics: $u^{i}_{1} = 16\xi^{i}\tau^2 + \ldots$, 
$u^{i}_{2} = \frac{512}{3}\xi^{i}\tau^4 + \ldots$, 
$u^{i}_{3} = \frac{69632}{45}\xi^{i}\tau^6 + \ldots$, $u^{i}_{4} = \frac{4063232}{315}\xi^{i}\tau^8 + \ldots$. 
The quantum phase flow does not satisfy the condition for 
canonicity: $\{u^{i}(\xi,\tau),u^{j}(\xi,\tau)\} = - I^{ij}(1 + 32 \hbar^2 \tau^2  + \frac{1792}{3} \hbar^4 \tau^4 + \ldots$). 
The Moyal bracket is consistent
with unitarity, $u^{i}(\xi,\tau) \wedge u^{j}(\xi,\tau) = - I^{ij}$, 
at least to order $O(\hbar^{16} \tau^{16})$.
}

\subsection{Green function for first-order quantum corrections in terms 
of Jacobi fields.}

Let us establish some useful identities for Jacobi fields. The energy
conservation implies 
\[
\mathcal{H}(\xi )\equiv \mathcal{H}(u_{0}(\xi,\tau )). 
\]
Taking derivative from this identity, one gets 
\begin{equation}
\frac{\partial \mathcal{H}(\xi )}{\partial \xi ^{s}}=\frac{\partial \mathcal{%
H}(u_{0}(\xi,\tau ))}{\partial u^{r}}J_{.s}^{r.}(\xi,\tau ).  \label{IDEN1}
\end{equation}
These are the six integrals of motion involving both the classical
trajectory and its rank-two Jacobi fields.

By writing the Hamilton equations in the form 
\[
\frac{\partial u_{0}^{i}(\xi,\tau )}{\partial \tau }=\{u_{0}^{i}(\xi,\tau
),\mathcal{H}(u_{0}(\xi,\tau ))\}=\{u_{0}^{i}(\xi,\tau ),\mathcal{H}(\xi
)\} 
\]
one gets 
\begin{equation}
\frac{\partial u_{0}^{i}(\xi,\tau )}{\partial \tau }=J_{.k}^{i.}(\xi,\tau
)(-I^{kl})\frac{\partial \mathcal{H}(\xi )}{\partial \xi ^{l}}.
\label{IDEN2}
\end{equation}

From canonicity of the classical hamiltonian flow one has 
\begin{equation}
J_{.r}^{i.}(\xi,\tau )I^{rs}J_{.s}^{k.}(\xi,\tau )=-\{u_{0}^{i}(\xi,\tau
),u_{0}^{k}(\xi,\tau )\}=I^{ik}.  \label{INVIJF}
\end{equation}

Equation (\ref{INVIJF}) implies that $-J_{r.}^{.k}(\xi,\tau )$ is the
inverse matrix of $J_{.r}^{k.}(\xi,\tau )$, and so 
\begin{eqnarray}
J_{.r}^{i.}(\xi,\tau )J_{k.}^{.r}(\xi,\tau ) &=&-\delta _{k}^{i},
\label{ORT1} \\
J_{.k}^{s.}(\xi,\tau )J_{s.}^{.i}(\xi,\tau ) &=&-\delta _{k}^{i}.
\label{ORT2}
\end{eqnarray}
Multiplying Eq.(\ref{IDEN2}) by $J_{i.}^{.r}(\xi,\tau )$ and performing the
summation over $i$ one gets Eq.(\ref{IDEN1}). These equations are therefore
not independent.

The evolution equation for the first-order quantum correction has the form 
\begin{equation}
\frac{\partial u_{1}^{i}(\xi,\tau )}{\partial \tau }=-I^{ik}\frac{\partial
^{2}\mathcal{H}(u_{0}(\xi,\tau ))}{\partial u_{0}^{k}\partial u_{0}^{l}}%
u_{1}^{l}(\xi,\tau )+J^{i}(u_{0}(\xi,\tau )), \label{EQMO}
\end{equation}
where $J^{i}(u_{0}(\xi,\tau ))$ is the inhomogeneous part of Eq.(\ref{FORCE}).
For periodic orbits, the system of equations for $u_{1}^{i}(\xi,\tau)$ looks like a system of equations for coupled oscillators whose frequencies are time-dependent. The functions $%
J^{i}(u(\xi,\tau ))$ play the role of an external time-dependent force.

We look for solutions of equation (\ref{EQMO}) in the form 
\begin{equation}
u_{1}^{i}(\xi,\tau )=C^{s}(\xi,\tau )J_{.s}^{i.}(\xi,\tau ).
\end{equation}
Taking equations of motion (7.1) for the rank-two Jacobi fields into account, one
gets 
\begin{equation}
J_{.s}^{i.}(\xi,\tau )\frac{\partial }{\partial \tau }C^{s}(\xi,\tau
)=J^{i}(u_{0}(\xi,\tau )).
\end{equation}
Using Eq.(\ref{ORT2}) and the initial conditions $u_{1}^{i}(0,\xi )=0,$ we
obtain 
\begin{equation}
u_{1}^{i}(\xi,\tau )=-J_{.s}^{i.}(\xi,\tau )\int_{0}^{\tau
}J_{r.}^{.s}(\tau ^{\prime },\xi )J^{r}(u_{0}(\tau ^{\prime },\xi ))d\tau
^{\prime }.  \label{SOLU1}
\end{equation}

The function 
\[
G_{s}^{r}(\tau ,\tau ^{\prime },\xi )=-J_{.m}^{r.}(\xi,\tau
)J_{s.}^{.m}(\tau ^{\prime },\xi )\theta (\tau -\tau ^{\prime }) 
\]
entering the right-hand side of Eq.(\ref{SOLU1}) satisfies equation 
\begin{equation}
\left( \delta _{r}^{p}\frac{\partial }{\partial \tau }+I^{pq}\frac{\partial
^{2}\mathcal{H}(u_{0}(\xi,\tau ))}{\partial u_{0}^{q}\partial u_{0}^{r}}%
\right) G_{s}^{r}(\tau ,\tau ^{\prime },\xi )=\delta _{s}^{p}\delta (\tau
-\tau ^{\prime })  \label{eqGF}
\end{equation}
and can be recognized as the Green function for the first-order quantum
correction to the classical trajectory.

\section{Quantum corrections to Kepler orbits}

The classical Kepler problem is described in many textbooks. This problem has 
the exact solution in quantum mechanics also (see e.g. \cite{LALIQM}). 
We construct the lowest order quantum corrections to Kepler orbits.

In order to calculate Jacobi fields, one has to keep the initial
canonical variables as free parameters. It means that we have to analyze
Kepler orbits in arbitrary coordinate system $K^{\prime}$. When all the derivatives
over $\xi$ are calculated, one can pass to a coordinate system $K$ where the orbits belong to a $(x,y)$ plane and where the expressions simplify.

\subsection{Spherical basis}

In the Cartesian coordinate system, the basis vectors are defined by 
\begin{eqnarray*}
\mathbf{e}_{x} &=&(1,0,0), \\
\mathbf{e}_{y} &=&(0,1,0), \\
\mathbf{e}_{z} &=&(0,0,1),
\end{eqnarray*}
The Kepler problem is usually treated in the spherical basis. The spherical
basis vectors parametrized by radius $r,$ polar angle $\theta$, and azimuthal angle $\varphi $ have the form
\begin{eqnarray}
\mathbf{e}_{r} &=&(\sin \theta \cos \varphi ,\sin \theta \sin \varphi ,\cos
\theta ),  \label{a} \\
\mathbf{e}_{\theta } &=&(\cos \theta \cos \varphi ,\cos \theta \sin \varphi
,-\sin \theta )=\frac{\partial \mathbf{e}_{r}}{\partial \theta },  \label{b}
\\
\mathbf{e}_{\varphi } &=&(-\sin \varphi ,\cos \varphi ,0)=\frac{1}{\sin
\theta }\frac{\partial \mathbf{e}_{r}}{\partial \varphi }.  \label{c}
\end{eqnarray}
They are orthonormal and obey the scalar and vector product rules:
\begin{eqnarray*}
\mathbf{e}_{i}\mathbf{e}_{j} &=&\delta _{ij}, \\
\mathbf{e}_{i}\times \mathbf{e}_{j} &=&\varepsilon _{ijk}\mathbf{e}_{k}
\end{eqnarray*}
for $i,j=r,\theta ,\varphi $.

In the spherical basis, the particle velocity can be decomposed as follows: 
\begin{eqnarray*}
\mathbf{\dot{x}} &=&\frac{\partial r\mathbf{e}_{r}}{\partial \tau}=\dot{r}%
\mathbf{e}_{r}+r\dot{\theta}\frac{\partial \mathbf{e}_{r}}{\partial \theta }%
+r\dot{\varphi}\frac{\partial \mathbf{e}_{r}}{\partial \varphi } \\
&=&\dot{r}\mathbf{e}_{r}+r\dot{\theta}\mathbf{e}_{\theta }+r\sin \theta 
\mathbf{e}_{\varphi }\dot{\varphi}.
\end{eqnarray*}

The Lagrangian of the system:
\[
L=\frac{m\mathbf{\dot{x}}^{2}}{2}=\frac{m}{2}(\dot{r}^{2}+r^{2}\dot{\theta}%
^{2}+r^{2}\sin ^{2}\theta \dot{\varphi}^{2})+\frac{\alpha }{r}. 
\]
The canonical momenta are as follows 
\begin{eqnarray}
p_{r} &=&m\dot{r},  \label{MOME1} \\
p_{\theta } &=&mr^{2}\dot{\theta},  \label{MOME2} \\
p_{\varphi } &=&mr^{2}\dot{\varphi}\sin ^{2}\theta .  \label{MOME3}
\end{eqnarray}

In the spherical basis, the canonical momenta have the form 
\begin{equation}
\mathbf{p}=m\mathbf{\dot{x}=}p_{r}\mathbf{e}_{r}+\frac{p_{\theta }}{r}%
\mathbf{e}_{\theta }+\frac{p_{\varphi }}{r\sin \theta }\mathbf{e}_{\varphi
}=\left( 
\begin{array}{l}
p_{r}\sin \theta \cos \varphi +\frac{p_{\theta }}{r}\cos \theta \cos \varphi
-\frac{p_{\varphi }}{r\sin \theta }\sin \varphi \\ 
p_{r}\sin \theta \sin \varphi +\frac{p_{\theta }}{r}\cos \theta \sin \varphi
+\frac{p_{\varphi }}{r\sin \theta }\cos \varphi \\ 
p_{r}\cos \theta -\frac{p_{\theta }}{r}\sin \theta
\end{array}
\right)  \label{MOME}
\end{equation}
The Hamiltonian function equals 
\begin{equation}
\mathcal{H}=\frac{p_{r}^{2}}{2m}+\frac{p_{\theta }^{2}}{2mr^{2}}+\frac{%
p_{\varphi }^{2}}{2mr^{2}\sin ^{2}\theta }-\frac{\alpha }{r}.  \label{HAMI}
\end{equation}

The orbital momentum $\mathbf{L=x\times p}$ can be represented as follows 
\begin{equation}
\mathbf{L}=p_{\theta }\mathbf{e}_{\varphi }-\frac{p_{\varphi }}{\sin \theta }%
\mathbf{e}_{\theta }.  \label{LS}
\end{equation}
The absolute value of the orbital momentum equals 
\[
L^{2}=p_{\theta }^{2}+\frac{p_{\varphi }^{2}}{\sin ^{2}\theta }\mathbf{.} 
\]

It is useful to parametrize $p_{\theta }$ and $p_{\varphi }$ as
follows:
\begin{eqnarray}
p_{\theta } &=&L\sin \chi ,  \label{PT} \\
p_{\varphi } &=&L\cos \chi \sin \theta .  \label{PF}
\end{eqnarray}

\subsection{Euler rotation}

We wish to pass over to a coordinate system $K^{\prime }$ in which the particle trajectory belongs to the $(x^{\prime},y^{\prime})$ plane, while the orbital momentum is directed along the $z^{\prime}$-axis. Furthermore, we require the perigelium of ellipse along which the particle moves be on the positive side of the $x^{\prime}$ axis.

In the initial coordinate system, $K$, the Cartesian components of the orbital momentum can be found to be 
\begin{eqnarray*}
L_{x}/L &=&-\sin \chi \sin \varphi -\cos \chi \cos \theta \cos \varphi , \\
L_{y}/L &=&\sin \chi \cos \varphi -\cos \chi \cos \theta \sin \varphi , \\
L_{z}/L &=&\cos \chi \sin \theta .
\end{eqnarray*}
The vector $\mathbf{L}$ determines plane $(x^{\prime},y^{\prime})$ 
orthogonal to $\mathbf{L}$. This plane intersects the plane $(x,y)$ of the coordinate system $K$. The line of the intersection is parallel to $\mathbf{e}_{z}\times \mathbf{L.}
$ It forms an angle, $\phi _{E}$, with the $x$-axis. The first Euler
rotation brings $\mathbf{e}_{x}\ $parallel to $\mathbf{e}_{z}\times \mathbf{L%
}$\textbf{.} In order to find $\phi _{E}$, we notice that $\cos \phi
_{E}\varpropto \mathbf{e}_{x}(\mathbf{e}_{z}\times \mathbf{L})=(\mathbf{e}%
_{x}\times \mathbf{e}_{z})\mathbf{L=}-L_{y}$, $\sin \phi _{E}\varpropto 
\mathbf{e}_{y}(\mathbf{e}_{z}\times \mathbf{L})=(\mathbf{e}_{y}\times 
\mathbf{e}_{z})\mathbf{L=L}_{x}$ and $(\mathbf{e}_{z}\times \mathbf{L}%
)_{i}=\varepsilon _{i3k}L_{k}=(-L_{y},L_{x},0)$, so that $(\mathbf{e}%
_{z}\times \mathbf{L})^{2}=L^{2}-L_{z}^{2}$. The second Euler rotation
brings $\mathbf{e}_{z}\ $parallel to $\mathbf{L}$. The first pair of the
Euler angles $(\phi _{E},\vartheta _{E})$ can therefore be found from
equations 
\begin{eqnarray*}
\cos \phi _{E} &=&\frac{-\sin \chi \cos \varphi +\cos \chi \cos \theta \sin
\varphi }{\sqrt{1-\cos ^{2}\chi \sin ^{2}\theta }}, \\
\sin \phi _{E} &=&\frac{-\sin \chi \sin \varphi -\cos \chi \cos \theta \cos
\varphi }{\sqrt{1-\cos ^{2}\chi \sin ^{2}\theta }}, \\
\cos \vartheta _{E} &=&\cos \chi \sin \theta , \\
\sin \vartheta _{E} &=&\sqrt{1-\cos ^{2}\chi \sin ^{2}\theta }.
\end{eqnarray*}

Suppose the third Euler rotation by the angle $\psi _{E}$ brings the $x$%
-axis parallel to the major half-axis of the ellipse in the direction 
\textit{towards} the perigelium. Let us compare the unit vector $\mathbf{n=e}_{r}$
in the two coordinate systems. The Cartesian components of $\mathbf{n}$ are 
related as follows 
\[
\left\| E\right\| \left( 
\begin{array}{l}
\sin \theta \cos \varphi \\ 
\sin \theta \sin \varphi \\ 
\cos \theta
\end{array}
\right) =\left( 
\begin{array}{l}
\cos \varphi _{p}^{\prime } \\ 
\sin \varphi _{p}^{\prime } \\ 
0
\end{array}
\right) , 
\]
where $\left\| E\right\| $ is the Euler rotation matrix: 
\[
\left\| E\right\| =\left( 
\begin{array}{lll}
\cos \psi _{E} & \sin \psi _{E} & 0 \\ 
-\sin \psi _{E} & \cos \psi _{E} & 0 \\ 
0 & 0 & 1
\end{array}
\right) \left( 
\begin{array}{lll}
1 & 0 & 0 \\ 
0 & \cos \vartheta _{E} & \sin \vartheta _{E} \\ 
0 & -\sin \vartheta _{E} & \cos \vartheta _{E}
\end{array}
\right) \left( 
\begin{array}{lll}
\cos \phi _{E} & \sin \phi _{E} & 0 \\ 
-\sin \phi _{E} & \cos \phi _{E} & 0 \\ 
0 & 0 & 1
\end{array}
\right) 
\]
and $\varphi _{p}$ is the phase set off, i.e., the angular distance from
the perigelium at start of the motion ($\tau =0$). We obtain 
\begin{eqnarray*}
\cos \varphi _{p}^{\prime } &=&\sin \theta \cos \varphi \cos \psi _{E}\cos
\phi _{E}-\sin \theta \cos \varphi \sin \psi _{E}\cos \vartheta _{E}\sin
\phi _{E} \\
&&+\sin \theta \sin \varphi \cos \psi _{E}\sin \phi _{E}+\sin \theta \sin
\varphi \sin \psi _{E}\cos \vartheta _{E}\cos \phi _{E} \\
&&+\sin \psi _{E}\sin \vartheta _{E}\cos \theta , \\
\sin \varphi _{p}^{\prime } &=&-\sin \theta \cos \varphi \sin \psi _{E}\cos
\phi _{E}-\sin \theta \cos \varphi \cos \psi _{E}\cos \vartheta _{E}\sin
\phi _{E} \\
&&-\sin \theta \sin \varphi \sin \psi _{E}\sin {\phi _{E}}+\sin \theta \sin
\varphi \cos \psi _{E}\cos \vartheta _{E}\cos \phi _{E} \\
&&+\cos \psi _{E}\sin \vartheta _{E}\cos \theta , \\
0 &=&\sin {\vartheta _{E}}\sin \theta \sin (\phi _{E}-\varphi )+\cos
\vartheta _{E}\cos \theta .
\end{eqnarray*}

The last equation is the identity. The first two ones give 
\begin{eqnarray}
\cos \varphi _{p}^{\prime } &=&-\frac{p_{\theta }}{\sqrt{L^{2}-p_{\varphi
}^{2}}}\sin \theta \cos \psi _{E}+\frac{L}{\sqrt{L^{2}-p_{\varphi }^{2}}}%
\cos \theta \sin \psi _{E},  \label{CFP} \\
\sin \varphi _{p}^{\prime } &=&\frac{p_{\theta }}{\sqrt{L^{2}-p_{\varphi
}^{2}}}\sin \theta \sin \psi _{E}+\frac{L}{\sqrt{L^{2}-p_{\varphi }^{2}}}%
\cos \theta \cos \psi _{E}.  \label{SFP}
\end{eqnarray}
These equations can be solved for $\psi _{E}$ to give 
\begin{eqnarray}
\cos \psi _{E} &=&\frac{\cos \theta \sin \varphi _{p}^{\prime }-\sin \chi
\sin \theta \cos \varphi _{p}^{\prime }}{\sqrt{1-\cos ^{2}\chi \sin
^{2}\theta }}, \label{PSIE1} \\
\sin \psi _{E} &=&\frac{\cos \theta \cos \varphi _{p}^{\prime }+\sin \chi
\sin \theta \sin \varphi _{p}^{\prime }}{\sqrt{1-\cos ^{2}\chi \sin
^{2}\theta }}. \label{PSIE2}
\end{eqnarray}

The canonical momenta in the coordinate system $K^{\prime }$ can be found to be 
\[
\left\| E\right\| \left( 
\begin{array}{l}
p_{x} \\ 
p_{y} \\ 
p_{z}
\end{array}
\right) =\left( 
\begin{array}{l}
p_{r}\cos \varphi _{p}^{\prime }-\frac{L}{r}\sin \varphi _{p}^{\prime } \\ 
p_{r}\sin \varphi _{p}^{\prime }+\frac{L}{r}\cos \varphi _{p}^{\prime } \\ 
0
\end{array}
\right) , 
\]
so we conclude (cf. \ref{MOME}) $p_{r}^{\prime }=p_{r}$, $p_{\theta
}^{\prime }=0$, and $p_{\varphi }^{\prime }=L.$ Of course, 
\[
\left\| E\right\| \left( 
\begin{array}{l}
L_{x} \\ 
L_{y} \\ 
L_{z}
\end{array}
\right) =\left( 
\begin{array}{l}
0 \\ 
0 \\ 
L
\end{array}
\right) . 
\]

\subsection{Orbits in the coordinate system K}

We use the atomic units $m = \alpha = \hbar =1$. The classical orbits in the
coordinate space are known to be the ellipses. The value $p$ (do mot mix
with momentum!), eccentricity $e$, major half-axis $a$ of the ellipse have the form (see e.g. \cite{LALICM}): 
\begin{eqnarray*}
p &=&L^{2}, \\
e &=&\sqrt{1+2EL^{2}}=\cos \chi . \\
a &=&\frac{p}{\sin ^{2}\chi }.
\end{eqnarray*}
Given these parameters are known as functions of the initial canonical
variables $\xi ^{i}=(r,\theta ,\varphi ,p_{r},p_{\theta },p_{\varphi })$, one
can find the set off angle $\varphi _{p}^{\prime }$ from equation
\begin{equation}
r=\frac{p}{1+\cos \chi \cos \varphi _{p}^{\prime }}.  \label{e9}
\end{equation}
The Euler angle $\psi _{E}$ is then fixed by Eqs.(\ref{PSIE1}) - (\ref{PSIE2}).
The Euler angles $(\phi _{E},\vartheta _{E},\psi _{E})$ are then determined.

In the coordinate system $K^{\prime },$ the classical orbit has the form 
\begin{eqnarray}
r_{\tau } &=& r_{\tau }^{\prime} = \frac{p}{1+\cos \chi \cos \varphi _{\tau }^{\prime }},
\label{e1} \\
r_{\tau } &=& r_{\tau }^{\prime} = a(1-\cos \chi \cos \nu ),  \label{e2} \\
\tau &=&p^{3/2}\frac{\nu -\cos \chi \sin \nu }{\sin ^{3}\chi },  \label{e3}
\\
\cos \varphi _{\tau }^{\prime } &=&\frac{\cos \nu -\cos \chi }{1-\cos \chi
\cos \nu },  \label{e4} \\
\sin \varphi _{\tau }^{\prime } &=&\frac{\sin \chi \sin \nu }{1-\cos \chi
\cos \nu },  \label{e5} \\
p_{r\tau } &=&p_{r\tau }^{\prime }=\frac{\frac{\partial r_{\tau }}{\partial
\nu }}{\frac{\partial \tau }{\partial \nu }},  \label{e6} \\
p_{\theta \tau }^{\prime } &=&0,  \label{e8} \\
p_{\varphi \tau }^{\prime } &=&\sqrt{p}.  \label{10}
\end{eqnarray}
The motion starts at $\nu _{p}$ which corresponds to $\tau _{p}$ Eq.(\ref{e3}%
) and $\varphi _{p}^{\prime }$ Eq.(\ref{e9}). The initial conditions are as follows: 
\[
(r^{\prime },\theta ^{\prime },\varphi ^{\prime },p_{r}^{\prime },p_{\theta
}^{\prime },p_{\varphi }^{\prime })_{\tau =\tau _{p}}=(r,\frac{\pi }{2}%
,\varphi _{p}^{\prime },p_{r},0,L), 
\]
the values $\varphi _{p}^{\prime }$ and $L=\sqrt{p}$ are functions of $\xi $.

Now, we are in a position to find the angular coordinates of the trajectory in the $K$ system:
\[
\left( 
\begin{array}{l}
\sin \theta _{\tau }\cos \varphi _{\tau } \\ 
\sin \theta _{\tau }\sin \varphi _{\tau } \\ 
\cos \theta _{\tau }
\end{array}
\right) =\left\| E^{-1}\right\| \left( 
\begin{array}{l}
\cos \varphi _{\tau }^{\prime } \\ 
\sin \varphi _{\tau }^{\prime } \\ 
0
\end{array}
\right) , 
\]
where $\left\| E^{-1}\right\| $ is the inverse Euler matrix. The radii in $K$ and $K^{\prime }$ coincide. The canonical momenta in $K$ may be found with the help of Eqs.(\ref{MOME1}) - (\ref{MOME3}).

We thus get the phase-space trajectories in the initial coordinate system: 
\[
u_{0}^{i}(\xi,\tau )=(r_{\tau },\theta _{\tau },\varphi _{\tau },p_{r\tau
},p_{\theta \tau },p_{\varphi \tau }). 
\]

\subsection{Jacobi fields in parametric representation}

The phase-space trajectories represented in parametric forms bring specific features in the calculation of the Jacobi fields. In our case, the orbits are parametrized by the parameter $\nu $ which is a function of time $%
\tau $ \textit{and} initial canonical variables $\xi ^{i}$. The derivatives
of the functions $u^{i}(\nu ,\xi )$ which we have constructed explicitly involve
the derivatives of $\nu =\nu (\xi,\tau ).$

Let $\tau _{p}(\xi )$ be time needed to come to $\xi $ from the perigelium ($%
x^{\prime }=r,$ $y^{\prime }=0$). We have 
\begin{equation}
\tau _{p}(\xi )=p^{3/2}\frac{\nu _{p}-\cos \chi \sin \nu _{p}}{\sin ^{3}\chi 
}.  \label{TAU8}
\end{equation}
The dependence on $\xi $ comes through the parameters $p$, $\chi $, and $\nu
_{p}.$ The parameter $\nu _{p}=\nu _{p}(\xi )$ is determined from Eqs.(\ref
{e9}), (\ref{e4}) and (\ref{e5}).

At $\tau >\tau _{p}$, we shift in Eq.(\ref{e3}) the time variable $\tau
\rightarrow \tau +\tau _{p}(\xi )$ and obtain 
\begin{equation}
\tau +\tau _{p}(\xi )=p^{3/2}\frac{\nu -\cos \chi \sin \nu }{\sin ^{3}\chi }%
\equiv \mathcal{T}(\nu ,\xi ).  \label{TAU}
\end{equation}
According to the convention, the motion starts at $\tau =0$. Now, the parameter $\tau $ is independent of $\xi ^{i}$. Taking the first and second derivatives
of Eq.(\ref{TAU}), we obtain 
\begin{eqnarray}
\frac{\partial \tau _{p}(\xi )}{\partial \xi ^{i}} &=&\frac{\partial 
\mathcal{T}(\nu ,\xi )}{\partial \xi ^{i}}+\frac{\partial \mathcal{T}(\nu
,\xi )}{\partial \nu }\frac{\partial \nu }{\partial \xi ^{i}}, \\
\frac{\partial ^{2}\tau _{p}(\xi )}{\partial \xi ^{i}\partial \xi ^{k}} &=&%
\frac{\partial ^{2}\mathcal{T}(\nu ,\xi )}{\partial \xi ^{i}\partial \xi ^{k}%
}+\frac{\partial ^{2}\mathcal{T}(\nu ,\xi )}{\partial \xi ^{i}\partial \nu }%
\frac{\partial \nu }{\partial \xi ^{k}}+\frac{\partial ^{2}\mathcal{T}(\nu
,\xi )}{\partial \xi ^{k}\partial \nu }\frac{\partial \nu }{\partial \xi ^{i}%
} \nonumber \\
&+&\frac{\partial ^{2}\mathcal{T}(\nu ,\xi )}{\partial \nu ^{2}}\frac{%
\partial \nu }{\partial \xi ^{k}}\frac{\partial \nu }{\partial \xi ^{i}}
+\frac{\partial \mathcal{T}(\nu ,\xi )}{\partial \nu }\frac{\partial ^{2}\nu 
}{\partial \xi ^{i}\partial \xi ^{k}}.
\end{eqnarray}
These equations allow to find the derivatives $\frac{\partial \nu }{\partial
\xi ^{i}}$ and $\frac{\partial ^{2}\nu }{\partial \xi ^{i}\partial \xi ^{k}}$
at fixed $\tau $. The dependence on $\tau $ enters implicitly through $\nu $.

Given the time dependence in the parametric form, one finds the total derivatives and the Jacobi fields in the parametric representation, accordingly: 
\begin{eqnarray}
J_{.i}^{l.}(\xi,\tau ) &=&\frac{\partial u^{l}(\nu ,\xi )}{\partial \xi ^{i}%
}+\frac{\partial u^{l}(\nu ,\xi )}{\partial \nu }\frac{\partial \nu }{%
\partial \xi ^{i}}, \\
J_{.ik}^{l..}(\xi,\tau ) &=&\frac{\partial ^{2}u^{l}(\nu ,\xi )}{\partial
\xi ^{i}\partial \xi ^{k}}+\frac{\partial ^{2}u^{l}(\nu ,\xi )}{\partial \xi
^{i}\partial \nu }\frac{\partial \nu }{\partial \xi ^{k}}+\frac{\partial
^{2}u^{l}(\nu ,\xi )}{\partial \xi ^{k}\partial \nu }\frac{\partial \nu }{%
\partial \xi ^{i}} \nonumber \\
&+&\frac{\partial ^{2}u^{l}(\nu ,\xi )}{\partial \nu ^{2}}%
\frac{\partial \nu }{\partial \xi ^{i}}\frac{\partial \nu }{\partial \xi ^{k}%
}+\frac{\partial u^{l}(\nu ,\xi )}{\partial \nu }\frac{\partial ^{2}\nu }{%
\partial \xi ^{i}\partial \xi ^{k}}.
\end{eqnarray}

The formulae simplify considerably if we set $K^{\prime }=K$ after taking the derivatives. The explicit parametric form of $J_{i.}^{.l}(\xi,\tau )$ and $%
J_{.ik}^{l..}(\xi,\tau )$ can be found using MAPLE. We give results for $J_{.i}^{l.}(\xi,\tau )$ for circular orbits only ($\eta =\pi /2$) where one can set $\nu _{p}=\varphi_{p}=0$: 
\begin{equation}
\left\| J_{.i}^{l.}(\xi,\tau )\right\| =\left| \stackrel{i\rightarrow }{
\begin{array}{cccccc}
\cos \nu & 0 & 0 & {p}^{3/2}\sin \nu & 0 & 2\,{p}^{1/2}\left( 1-\cos \nu
\right) \\ 
0 & \cos \nu & 0 & 0 & {p}^{-1/2}\sin \nu & 0 \\ 
-2\,{p}^{-1}\sin \nu & 0 & 1 & 2{p}^{1/2}\,\left( \cos \nu -1\right) & 0 & {p%
}^{-1/2}(4\,\sin \nu -3\,\nu ) \\ 
-{p}^{-3/2}\sin \nu & 0 & 0 & \cos \nu & 0 & 2\,{p}^{-1}\sin \nu \\ 
0 & -{p}^{1/2}\sin \nu & 0 & 0 & \cos \nu & 0 \\ 
0 & 0 & 0 & 0 & 0 & 1
\end{array}
{}}\right|  \label{CIRCJF}
\end{equation}
The rank-two and rank-three Jacobi fields which we have constructed satisfy for arbitrary values $p$ and $\chi $ the initial conditions $%
J_{.i}^{l.}(0,\xi )=\delta _{i}^{l}$ and $J_{.ik}^{l..}(0,\xi )=0$, equations of motion (\ref{GRADI}), the conservation laws Eq.(\ref{IDEN2}), and the canonicity condition (\ref{INVIJF}).

\vspace{8mm} 
\begin{figure}[!htb]
\begin{center}
\includegraphics[angle=0,width=6.5 cm]{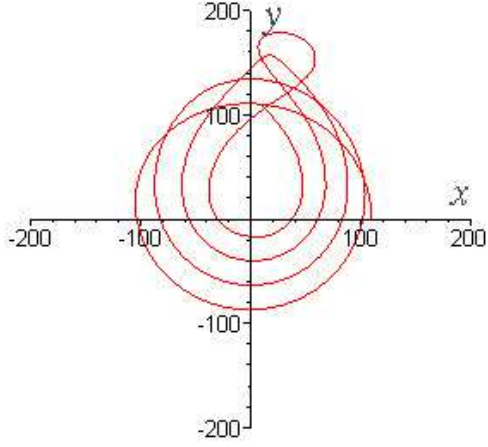}
\end{center}
\caption{The lowest order quantum Kepler orbit coresponding to the classical circular orbit $r=p=110$ in the atomic units $m = \alpha = \hbar = 1$ for $0 \leq \nu  \leq 8\pi$. The motion starts at $x=r$ and $y=0$.
}
\label{fig9}
\end{figure}

\subsection{Quantum orbits and resonance}

In general case, the functions $u_{1}^{i}(\xi,\tau )$ are expressed as one-dimensional integrals. Here, we report results for circular orbits, where significant simplifications occur.

The external force is found to be 
\begin{eqnarray*}
J^{3} &=&p^{-5/2}(-1+\cos \nu )(5\cos ^{2}\nu -7\cos \nu +1), \\
J^{4} &=&3p^{-3}(-1+\cos \nu )(2\cos ^{2}\nu -2\cos \nu -1),
\end{eqnarray*}
while $J^{1}=J^{2}=J^{5}=J^{6}=0.$ The components $i=1,2,5,$ and $6$ vanish
for all orbits ($0<\eta <0$). Also, for all orbits $J^{3}\sim p^{-5/2}$ and $%
J^{4}\sim p^{-3}$.

The external force is periodic. The matrix entering the left-hand side of Eq.(\ref{eqGF}) for circular orbits has the form

\[
\left\| -I^{ir}\frac{\partial ^{2}\mathcal{H}(u_{0}(\xi,\tau ))}{\partial
u^{r}\partial u^{k}}\right\| =\stackrel{k\rightarrow }{\left| 
\begin{array}{cccccc}
0 & 0 & 0 & 1 & 0 & 0 \\ 
0 & 0 & 0 & 0 & p^{-2} & 0 \\ 
-2p^{-5/2} & 0 & 0 & 0 & 0 & p^{-2} \\ 
-p^{-3} & 0 & 0 & 0 & 0 & 2\,p^{-5/2} \\ 
0 & -p^{-1} & 0 & 0 & 0 & 0 \\ 
0 & 0 & 0 & 0 & 0 & 0
\end{array}
\right| }.
\]
Inspecting this matrix, we find that frequency of the external force coincides with the eigenfrequency of effective oscillator. This condition is sufficient for the resonance.

The Green function can be constructed using Eq.(\ref{CIRCJF}). The first-order corrections become 
\begin{eqnarray*}
u_{1}^{1} &=&-\frac{3}{4}\cos ^{3}\nu +4\cos ^{2}\nu +\frac{7}{4}\cos \nu -5+%
\frac{15}{4}\nu \sin \nu , \\
u_{1}^{3} &=&p^{-1}\left( -10\sin \nu \cos \nu +\frac{4}{3}\sin \nu +\frac{13%
}{6}\sin \nu \cos ^{2}\nu +\frac{15}{2}\nu \cos \nu -\nu \right) , \\
u_{1}^{4} &=&p^{-3/2}\left( \frac{9}{4}\cos ^{2}\nu \sin \nu -8\cos \nu \sin
\nu +2\sin \nu +\frac{15}{4}\nu \cos \nu \right) ,
\end{eqnarray*}
while other components vanish: $u_{1}^{2}=u_{1}^{5}=u_{1}^{6}=0.$ The
corrections $u_{1}^{2},$ $u_{1}^{5},$ and $u_{1}^{6}$ vanish for $0<\eta <0$ also. The observed dependence of $u_{1}^{1},$ $u_{1}^{3},$ and $u_{1}^{4}$ on $p$ holds for arbitrary $\eta $ and $\varphi _{p}^{\prime }.$

The amplitude of the quantum corrections increases linearly with time, which is the feature inherent to the resonances.

In Fig. \ref{fig9}, we plot the radius $r_{\tau }=r+r_{1}$ versus the azimuthal angle $\varphi _{\tau }=\varphi _{0}+\varphi _{1}$ in the polar coordinate system for $p=110$ during the four $\nu$-periods ($0\leq \nu \leq 8\pi $), the motion starts at $\nu =\varphi _{0}=0.$ The value of $p$ is taken to be large in order to fulfill the semiclassical conditions where the first quantum correction is supposed to be small. The region of validity of the solutions
is restricted to $|u_{1}^{i}(\xi,\tau )|\ll |u_{0}^{i}(\xi,\tau )|.$ We see from the plot, that for $p=110$ one-two periods is the upper admissible limit to which the first-order quantum characteristics can be extrapolated. This is in the obvious contrast with the operator methods where the semiclassical approximation usually starts to work from very low quantum numbers.

\section{Many-body potential scattering}
\setcounter{equation}{0}

Suppose we have an $n$-body system which evolves quantum-mechanically due to an interaction through a potential. The initial-state wave function is assumed to be known, so we can construct the initial-state Wigner function. 

\subsection{Calculation of Jacobi fields using bunches of characteristics}

The classical trajectories $u^{i}_{0}(\xi,\tau)$ are assumed to be constructed  numerically. In order to evaluate Jacobi fields $\partial^{t} u^{i}_{0}(\xi,\tau)/\partial \xi^{i_{1}} ... \partial \xi^{i_{t}}$ to order $O(\hbar^2)$, one may consider a bunch of characteristics coming out from a neighborhood of $\xi^{i}$. Using the simplest finite difference method, $3^{2n}$ such characteristics should be calculated and stored.

The order $O(\hbar^4)$ involves Jacobi fields of $u^{i}_{0}(\xi,\tau)$ with $t=1,2,3,4$ and Jacobi fields of $u^{i}_{1}(\xi,\tau)$ with $t=1,2$. Accordingly, one has to construct a bunch of at least $5^{2n}$ characteristics $u^{i}_{0}(\zeta,\tau)$ and at least $3^{2n}$ characteristics $u^{i}_{1}(\zeta,\tau)$ where $\zeta^{i}$'s belong to a neighborhood of $\xi^{i}$. 

The appearance of a bunch of characteristics is a manifestation of non-locality in quantum mechanics. Moving up with the Planck's constant expansion, one has to generate increasing bunches of characteristics. 
The number of characteristics increases exponentially. In the potential scattering of two particles, we would have to store $\sim 10^6$ characteristics, whereas in the potential scattering of four particles, we would already have to store $\sim 10^{11}$ characteristics to order $\hbar^2$.

\subsection{Calculation of Jacobi fields using ODE for Jacobi fields}

It is possible to reduce the amount of data generated and stored by computer. 
The main problem is to evaluate Jacobi fields.

Remarkably, Jacobi fields can be constructed 
by solving a system of ODE:
\begin{eqnarray}
\frac{\partial }{\partial \tau } J^{i.}_{.j}(\xi,\tau)  &=& F^{i}(u_{0})_{,k}  J^{k.}_{.j}(\xi,\tau),  \nonumber \\
\frac{\partial }{\partial \tau }J^{i..}_{.jk}(\xi,\tau) &=& F^{i}(u_{0})_{,lm} J^{l.}_{.j}(\xi,\tau) J^{m.}_{.k}(\xi,\tau) +
F^{i}(u_{0})_{,l} J^{l..}_{.jk}(\xi,\tau)
\label{GRADI}
\end{eqnarray}
with initial conditions
\begin{eqnarray}
J^{i.}_{.j}(\xi,0) &=& \delta^{i}_{j}, \nonumber \\
J^{i..}_{jk}(\xi,0) &=& 0.
\label{IGRADI}
\end{eqnarray}
Equations (\ref{GRADI}) follow from (\ref{RECU}) for $s=0$. Equations (\ref{GRADI}) 
are sufficient for constructing characteristics to order $O(\hbar^2)$. To find 
the first correction 
$u^{i}_{1}(\xi,\tau)$, it is necessary to calculate $(2n)^2$ functions 
$\partial u^{i}_{0}(\xi,\tau)/\partial \xi^{j}$ and $(2n)^{2}(2n+1)/2$ functions $\partial^2 u^{i}_{0}(\xi,\tau)/\partial \xi^{j}\partial \xi^{k}$. 
One has to store, therefore, $2n(n + 1)(2n + 1)$ functions describing characteristics and their derivatives instead of $2n \times 3^{2n}$ ones describing a bunch of characteristics.

The method of the previous subsection requires computer time of about 
$\tau_c  \sim 2n(2s + 1)^{2n}$, whereas the present method gives 
$\tau_c  \sim (2n)^{2s + 1}$ to order $O(\hbar^{2s})$. The number of characteristics and Jacobi fields in the present method grows exponentially with $s$. It means that for a few-body system, a high order $\hbar$ calculation is more effective in terms of bunches of characteristics, whereas for a many-body scattering and a low $s$ the method based on the propagation of Jacobi fields is preferable.

To order $O(\hbar^4)$, one has to construct derivatives of 
$u^{i}_{0}(\xi,\tau)$ up to the fourth degree. We thus have 
to store $2n$ functions $u^{i}_{0}(\xi,\tau)$ and 
$2n\sum_{s=1}^{4}(s + 2n - 1)!/(s!(2n - 1)!)$ Jacobi fields. 
The corresponding equations can be obtained by taking derivatives of the classical Hamilton's equations with respect to initial values of the canonical variables. The first and second derivatives of $u^{i}_{1}(\xi,\tau)$ enter the problem also. The corresponding equations can be obtained by taking derivatives of Eq.(\ref{RECU}) for $s = 1$.

The recursion is consistent: Jacobi fields $\partial^{t} u^{i}_{s}(\xi,\tau)/\partial \xi^{i_{1}}...\partial \xi^{i_{t}}$ 
are calculated sequentially towards increasing orders $s$ and $t$. 
The derivatives of Eq.(\ref{RECU}) involve derivatives of 
$u^{i}_{r}(\xi,\tau)$ with $r \leq s$ only.

In the potential scattering of four particles ($n=12$), we would have to 
calculate 24, 7800, and 499200 functions describing characteristics and 
Jacobi fields to orders $\hbar^0$, $\hbar^2$, and $\hbar^4$, respectively. As 
compared to the previous method, we gain a reduction in computer time by 
several orders of magnitude. In the potential scattering of two oxygen 
atoms ${^{16}O} + {^{16}O}$ (16 electrons and two nuclei, $n=54$ degrees 
of freedom), one has to calculate 108, 647460, and $\sim 10^{10}$ functions 
of time describing characteristics and Jacobi fields to orders $\hbar^0$, $\hbar^2$, 
and $\hbar^4$, respectively. The second order calculation might numerically 
be feasible.

\subsection{Monte Carlo method for averaging over the Wigner function}

The average value of an observable associated to an operator $\mathfrak{f}$ 
can be calculated
using equation $f(\xi,\tau) = f(\star u(\xi,\tau),0)$, expansion (\ref{FINA2}), 
and the initial-state Wigner function $W(\xi,0) \equiv W(\xi)$:
\begin{equation}
<f(\xi,\tau)> = \int \frac{d^{2n}\xi}{(2\pi \hbar)^{n}}f(\star u(\xi,\tau))W(\xi).
\label{AVER}
\end{equation}
We describe calculation of (\ref{AVER}) using an extension of the Monte Carlo method exploited in lattice QCD \cite{MART}. 

Let  us separate phase space into two regions $\Omega_{+}$ and $\Omega_{-}$ where 
the Wigner function is positive and negative definite, respectively. One has 
$W(\xi) = W_{+}(\xi) - W_{-}(\xi)$ where $W_{\pm}(\xi) \geq 0$.
The next step deals with generating events in $\Omega_{\pm}$ distributed 
according to probability densities $W_{\pm}(\xi)$. 

Consider region $\Omega_{+}$. One starts from generating $2n + 1$ numbers 
$(\xi^{i},\gamma)$ with $\xi^{i}$ distributed homogeneously in 
$\Omega_{+}$ and $\gamma$ distributed homogeneously in the 
interval $(0,W_{max})$ where 
\begin{equation}
W_{max} = \max_{\xi}W_{+}(\xi).
\end{equation}
The joint probability density of the $2n+1$-dimensional variable 
$(\xi^{i},\gamma)$ is assumed to be proportional to $\theta(W_{+}(\xi) - \gamma)$.
One can check that, owing to a constant normalization factor, the marginal 
probability density of $\xi^{i}$ is given by
\begin{equation}
W_{+}(\xi) = \int_{0}^{W_{max}}\theta(W_{+}(\xi) - \gamma)d\gamma.
\label{FANTA}
\end{equation}
Given $(\xi^{i},\gamma)$, we examine condition
\begin{equation}
W_{+}(\xi) - \gamma \geq 0. 
\label{CHECK}
\end{equation}
If it is fulfilled, we shift the number $N_{+}$ of successful events by one 
unit, set $\xi_{+ \; a} = \xi$, calculate quantum characteristic 
$u^{i}(\xi_{+ \; a},\tau)$, observable $f(\star u(\xi_{+ \; a},\tau))$, 
and keep the record with $a = N_{+}$. The star-product from the arguments 
of $f(\star u(\xi_{+ \; a},\tau))$ can be removed as described in Sect. IV. 
If inequality (\ref{CHECK}) is not fulfilled, next $2n + 1$ numbers $(\xi^{i},\gamma)$ 
are generated. 

The similar procedure applies for $\Omega_{-}$. 

Suppose we have generated samples of the $N_{+}$ and $N_{-}$ successful events 
$\xi_{\pm \;a} \in \Omega_{\pm}$. To get the average value of function $f(\xi,\tau)$, 
the values $f(\star u(\xi_{\pm\;a},\tau))$ should be multiplied by 
\begin{equation}
W_{\pm} = \int \frac{d^{2n}\xi}{(2\pi \hbar)^{n}}W_{\pm}(\xi),
\end{equation}
where $W_{+} - W_{-} = 1$, divided to total numbers $N_{\pm}$ of successful events, 
and summed up to give
\begin{equation}
<f(\xi,\tau)> \approx \frac{W_{+}}{N_{+}}\sum_{a=1}^{N_{+}}f(\star u(\xi_{+\;a},\tau)) - \frac{W_{-}}{N_{-}}\sum_{a=1}^{N_{-}}f(\star u(\xi_{-\;a},\tau)).
\label{AVERA}
\end{equation}

This equation accomplishes reduction of the evolution problem of an $n$-body  
quantum-mechanical system to a statistical-mechanical problem of computing an 
ensemble of quantum characteristics $u^{i}(\xi_{\pm \;a},\tau)$ and gradients 
$\partial^{t} u^{i}(\xi_{\pm \;a},\tau)/\partial \xi^{i_{1}} ... \partial \xi^{i_{t}} $.

The Monte-Carlo method is efficient for calculation of multidimensional integrals
of real positive functions. In quantum mechanics, the amplitudes are complex functions 
which oscillate rapidly in the semiclassical regime. The numerical calculation 
of oscillating functions requires the high precision, diminishing therefore the value 
of numerical methods for calculation of path integrals in the configuration space. The imaginary time 
formalism combined with the Monte-Carlo method is effective for finding binding energies of the 
lowest states, but not for scattering problems. The path-integral method in the 
phase space \cite{BLEAF1,MARIN} represents perhaps more promising tool, since the Green function 
is a real function, it is positive definite in the classical limit, although it is not positive 
definite in general. The phase-space path integral method is used in Ref. \cite{WONG} to find 
the evolution of a quantum state as an illustrative example (see also \cite{KRMAFU}).

\subsection{Scattering theory in terms the Wigner function}

The scattering theory treats elementary and bound particles essentially on
the same footing. We fix \textit{in} an \textit{out} scattering states at $%
\tau =\tau ^{\prime }$ and $\tau =\tau ^{\prime \prime },$ respectively, and
select clusters $\alpha $ and $\beta $ for elementary and bound particles
with definite momenta $\mathbf{p}_{\alpha }^{\prime }$ and $\mathbf{p}%
_{\beta }^{\prime \prime }$  
\begin{eqnarray}
W_{in}(\xi ) &=&\prod_{\alpha }\frac{1}{V}(2\pi \hbar )^{3}\delta (\mathbf{p}%
_{\alpha }^{\prime }-\sum_{i\in \alpha }\mathbf{p}_{i})W_{\alpha }(\xi
_{\alpha }),  \label{W2} \\
W_{out}(\xi ) &=&\prod_{\beta }\frac{1}{V}(2\pi \hbar )^{3}\delta (\mathbf{p}%
_{\beta }^{\prime \prime }-\sum_{i\in \beta }\mathbf{p}_{i})W_{\beta }(\xi
_{\beta }).
\end{eqnarray}
Here, $\xi =(\mathbf{x}_{1},...,\mathbf{x}_{N},\mathbf{p}_{1},...,\mathbf{p}%
_{N}),$ $N$ is the total number of particles participating in the reaction,
the number of clusters is less or equal to $N$, $V$ is the normalization
volume. The Wigner functions $W_{in}(\xi )$ and $W_{out}(\xi )$ correspond
to products of free plane waves of elementary particles and bound states
and include the product of the intrinsic Wigner functions of bound states.
We thus have $W_{\beta }(\xi _{\beta })=W_{\alpha }(\xi _{\alpha })=1$ for
elementary particles, whereas for bound states $W_{\beta }(\xi _{\beta })$
and $W_{\alpha }(\xi _{\alpha })$ represent the Wigner functions of bound
states in the rest frame, which are supposed to be known. The transition
probability from the initial \textit{in}-state in to the final \textit{out}%
-state is the absolute square of the $S$-matrix element $%
w_{fi}=|<out|in>|^{2}.$ In terms of the Wigner function, one finds 
\begin{equation}
w_{fi}=\int \frac{d^{2n}\xi }{(2\pi \hbar )^{n}}W_{out}(\star u(\xi ,\tau
))W_{in}(\xi ),  \label{CS}
\end{equation}
where $n=3N$. At this stage, Eq.(7.8) can be used for numerical simulations.

Equation (\ref{CS}) simplifies for scattering of one particle in an external
potential. The cross section is defined by 
\[
d\sigma _{fi} =\frac{1}{j_{inc}}\lim_{\tau ^{\prime \prime }\rightarrow +\infty
}\lim_{\tau ^{\prime }\rightarrow -\infty }\frac{w_{fi}}{\tau ^{\prime
\prime }-\tau ^{\prime }}\frac{Vd\mathbf{p}^{\prime \prime }}{(2\pi \hbar
)^{3}}.
\]
\newline
where $j_{inc}=\frac{v^{\prime }}{V}.$ In the classical limit 
\[
w_{fi}^{cl}=\frac{1}{V^{2}}\int d\mathbf{x}(2\pi \hbar )^{3}\delta
^{3}(p^{\prime \prime i}-u_{0}^{3+i}(\mathbf{x,p}^{\prime },\tau ))
\]
The initial momentum $\mathbf{p}^{\prime }$ is directed along the $z$-axis.
The scattering takes place on a finite-range potential. The limit for the
integral along the particle trajectory extends from $z=0$ to $z=v^{\prime
}(\tau ^{\prime \prime }-\tau ^{\prime })\rightarrow +\infty $ owing to
contributions $O(1)$ coming from the interaction region, which can be
neglected. The integrand does not depend on $z$, so this contribution is
factorized to give simply $v^{\prime }(\tau ^{\prime \prime }-\tau ^{\prime
}).$ As a second step, the change of variables $(\phi ,b)\rightarrow (\phi
,\theta )$ can be done, where $\phi $ is the azimuthal angle, $b$ is the
impact parameter and $\theta $ is the scattering polar angle. In this way,
one gets the well known expression 
\[
d\sigma _{fi}^{cl}=\frac{b}{\sin \chi }\frac{\partial b}{\partial \chi }d\Omega _{%
\mathbf{p}^{\prime \prime }}.
\]

The semiclassical expansion of $W_{out}(\star u(\xi ,\tau ))$ around $%
W_{out}(u(\xi ,\tau ))$ can be useful to calculate quantum corrections to
the classical cross sections.

\section{Conclusions}
\setcounter{equation}{0}

In this work, we discussed properties of the Weyl's symbols of the Heisenberg operators of 
canonical coordinates and momenta. Functions $u^{i}(\xi,\tau)$ which show quantum phase flow 
obey Hamilton's equations in quantum form (\ref{QF}). We derived these 
equations using the Green function method.
The principle of stationary action for quantum Hamilton's equations was formulated and 
quantum-mechanical extensions of the
Liouville theorem on conservation of the phase-space volume and the Poincar\'e theorem on conservation 
of $2p$-forms were found.

Symbols of generic Heisenberg operators are entirely determined by $u^{i}(\xi,\tau)$ from
\[
f(\xi,\tau) = f(\star u(\xi,\tau),0). 
\]
This equation is remarkable in many respects. It shows that quantum 
phase flow comprises the entire information on quantum evolution of symbols of Heisenberg 
operators. Usually characteristics 
satisfy first-order ODE e.g. the classical Hamilton's 
equations and solve first-order PDE e.g. the classical 
Liouville equation. Functions $u^{i}(\xi,\tau)$ are the genuine 
characteristics of Eq.(\ref{EVOL}), despite both 
$u^{i}(\xi,\tau)$ and $f(\xi,\tau)$ obey infinite-order PDE. It is quite surprising that to any fixed order of 
the semiclassical expansion $u^{i}(\xi,\tau)$ and therefore $f(\xi,\tau)$ can be constructed by solving a system 
of first-order ODE only. Such a circumstance allows to approach the 
problem numerically using efficient ODE integrators.

We discussed methods of eliminating the $\star$-symbol from arguments of functions $ f(\star u(\xi,\tau),0)$ by 
semiclassical expansion around $ f(u(\xi,\tau),0)$ and provided the explicit expressions up to the fourth order 
in $\hbar$.

The energy conservation in quantum form (\ref{EC}) renders upon expansion over the Planck's constant 
an infinite sequence of conserved quantities Eq.(\ref{ECHB}) depending on $u^{i}_{s}(\xi,\tau)$ and 
finite-order derivatives of $u^{i}_{s}(\xi,\tau)$ with respect to $\xi^{i}$ i.e. Jacobi fields. 
The order $\hbar^2$ conserved quantity (\ref{ECHB3})
is expressed in terms of classical trajectories and their Jacobi fields only.

For illustration purposes, we constructed the lowest order quantum corrections to 
the Kepler periodic circular orbits. These corrections increase linearly with time and 
show the resonance behavior.

The phase-space formulation of quantum dynamics can be 
useful for transport models in quantum chemistry and heavy-ion 
collisions where embedding coherence in propagation of 
particles remains, in particular, an unsolved problem. The principal 
advantage of the deformation quantization is its proximity 
to the classical picture of phase-space dynamics and the
strict quantum character concurrently. Specific effects 
such as quantum coherence and non-locality are displayed 
implicitly upon the $\hbar$ expansion in the dependence 
of physical quantities on Jacobi fields.

The main result of this work is the reduction by semiclassical expansion of 
quantum-mechanical evolution problem to a statistical-mechanical problem of 
computing an ensemble of quantum characteristics and their Jacobi fields. The
reduction to a system of ODE pertains rigorously at any fixed order in $\hbar$. 
Given quantum characteristics are constructed, physical observables can  
be found without further addressing to dynamics. The method of quantum characteristics 
can be useful for calculation of potential scattering of complex quantum systems. 

\vspace{5mm}
This work is supported in part by DFG grant No. 436 RUS 113/721/0-2, RFBR grant No. 06-02-04004, 
and European Graduiertenkolleg GR683.




\end{document}